\documentclass[fleqn,10pt]{wlscirep}
\usepackage[utf8]{inputenc}
\usepackage[T1]{fontenc}
\usepackage{amsfonts,amsmath,bbm,wasysym}
\usepackage{amsfonts,dsfont}
\usepackage{braket} 
\title{R\'enyi relative entropy based monogamy of entanglement in tripartite systems}
\author[1]{Marwa Manna\"i}
\author[1,2,3]{Hisham Sati}
\author[1,4,5,6]{Tim Byrnes}
\author[7,+]{Chandrashekar Radhakrishnan}
\affil[1]{Center for Quantum and Topological Systems, NYUAD Research Institute, New York University Abu Dhabi, UAE}
\affil[2]{Mathematics, Division of Science, New York University Abu Dhabi (NYUAD), Abu Dhabi, UAE}
\affil[3]{The Courant Institute for Mathematical Sciences, NYU, NY 10012, USA}
\affil[4]{New York University Shanghai, NYU- ECNU Institute of Physics at NYU Shanghai, 567 West Yangsi Road, 200124, China}
\affil[5]{State Key Laboratory of Precision Spectroscopy, School of Physical and Material Sciences, East China Normal University, Shanghai 200062, China}
\affil[6]{Department of Physics, New York University, New York, NY 10003, USA}
\affil[7]{Department of Computer Science and Engineering, New York University Shanghai, 567 West Yangsi Road, Pudong, Shanghai 200124, China}

\affil[+]{chandrashekar10@gmail.com}
\date{\today}

\begin{abstract}
A comprehensive investigation of the entanglement characteristics is carried out on tripartite spin-1/2 systems, examining prototypical tripartite states,
the thermal Heisenberg model, and the transverse field Ising 
model. The entanglement is computed using the R\'enyi relative entropy. In the traditional R\'enyi relative entropy,
the generalization parameter $\alpha$ can take values only in the range $0 \leq \alpha \leq 2$ due to the requirements of joint convexity of the measure.  To 
use the R\'enyi relative entropy over a wider range of $\alpha$, we use the sandwiched form which is jointly convex in the regime 
$0.5 \leq \alpha \leq \infty$. In prototypical tripartite states, we find that GHZ states are monogamous, but surprisingly so are W states.  
On the other hand, star states exhibit polygamy, due to the higher level of purity of the bipartite subsystems.  For spin models, we study the dependence 
of entanglement on various parameters such as temperature, spin-spin interaction, and anisotropy, and identify regions where entanglement is the largest.  
The R\'enyi parameter $\alpha$ scales the amount of entanglement in the system.  The entanglement measure based on the traditional and the sandwiched R\'enyi 
relative entropies obey the Araki-Lieb-Thirring inequality. 
In the Heisenberg models, namely the XYZ, XXZ, and XY models, the system is always monogamous. However, in the
transverse field Ising model, the state is initially polygamous and becomes monogamous with temperature and coupling.   
\end{abstract}

\begin{document}
\flushbottom
\maketitle
\thispagestyle{empty}

\section{Introduction}
\hspace{1.5em} Entanglement is a non-local quantum correlation phenomenon shared among multipartite quantum systems. Due to its potential applications in 
 cryptography \cite{cryptography1,cryptography2,cryptography3,cryptography4}, quantum teleportation \cite{telep1,telep2,telep3}, and quantum 
communication \cite{comm1,comm2,comm3}, entanglement is recognized as a resource required to perform quantum information processing. A natural task is 
then to quantify the amount of entanglement. Entanglement in bipartite systems is rather well-understood, and several sufficient conditions 
\cite{criteria1,criteria2,criteria3} have been proposed to detect whether a given quantum state is entangled or separable. For pure states, the von Neumann 
entanglement entropy is the de facto choice for measuring the amount of entanglement across a bipartition \cite{vonN}. For mixed states, concurrence 
is one of the most widely used measures to quantify entanglement in two-qubit systems \cite{concur1,concur2,concur3}. Other common approaches to 
compute entanglement are the logarithmic negativity \cite{neagtivity1,neagtivity2} and the relative entropy of entanglement \cite{relative}.  Of these 
measures, the relative entropy of entanglement can also be used to study multipartite systems. Several studies have been conducted to detect entanglement in multipartite quantum many-body systems.  In particular, it has been pointed out that apart from the bipartite entanglement, we can also use multipartite 
entanglement as an indicator of quantum phase transition
\cite{phasetransition1,phasetransition2,phasetransition3,phasetransition4,phasetransition5,phasetransition6}.  Further, it also known that multipartite 
entanglement has applications in quantum secret sharing \cite{secretsharing}, network key distribution \cite{keydistri}, high-precision quantum metrology \
\cite{metrology}, and multiparty quantum networks \cite{quanetwork1,quanetwork2}.

An important feature of entanglement is its distribution in multipartite systems.  This has been considered by Coffman, Kundu and Wootters through a study 
of the notion of monogamy of entanglement in tripartite systems \cite{monogamy2}.  This concept emphasizes that the maximal entanglement between two qubits 
$A$ and $B$, constrains their ability to share entanglement freely with a third qubit $C$.  Thus, the study of monogamy relation allows one to investigate 
the distribution of entanglement between different parts of the quantum system \cite{monogamy2}.  Monogamy is an essential feature that gives significant
advantages in achieving security in quantum key distribution protocols \cite{monogaQKDprotocol}. It also provides an effective tool for understanding 
various effects in quantum physics \cite{monogapp1,monogapp2,radhakrishnan2019time}, statistical mechanics \cite{statmonog}, and spin chain models \cite{condmatt1}.  
For example in Ref.\cite{monogamyXXZ} it has been shown that the monogamy inequality can identify quantum phase transitions in the Heisenberg XXZ model. 
Monogamy inequalities have been proposed for several different measures of entanglement such as concurrence \cite{monogamy2}, generalized concurrence 
\cite{generconcu1,generconcu2,generconcu3}, tangle \cite{tangle}, $n$-tangle \cite{ntangle}, and negativity \cite{negativmulti1,negativmulti2}.  However, 
these measures can only be used to study pure states.  For the case of mixed states, we must use monogamy relations based on a suitable measure that is 
applicable to mixed states, such as the relative entropy of entanglement.

Recently, entanglement measures based on both the Tsallis and R\'enyi quantum entropies, parameterized by a real number 
$\alpha$ have attracted significant attention \cite{Tsali,Tsali1,Tsali2,Tsali3,Tsali4,Tsali5}. 
In this work, we use the $\alpha$-R\'enyi relative entropy of entanglement and quantify the total entanglement in the multipartite system.  To compute 
entanglement over the complete range of $\alpha$, we use both the traditional R\'enyi relative entropy and the sandwiched R\'enyi relative entropy.
Individually, these two forms cannot cover the entire range of the generalization parameter $\alpha$.  Using these measures we evaluate the 
entanglement and monogamy relations for various prototypical tripartite systems and spin models.  For the spin-1/2 Heisenberg models,
we examine the {\it(i)} XYZ,  {\it(ii)} XXZ,  and {\it(iii)} XY chains.  We also consider the {\it(iv)} transverse-field Ising model. 
An investigation of the effect on entanglement of parameters such as temperature, 
spin-spin coupling, the anisotropy parameters, and the external magnetic field is examined in detail.  We study, in particular,
monogamy relations and identify the types of quantum states that result in monogamous and polygamous behavior.

The paper is arranged as follows. In Sec. \ref{measures}, we briefly review the definition of $\alpha$-R\'enyi entropy and also the relative 
entropy based entanglement measure defined using R\'enyi entropy.  A study of the monogamy relations of the GHZ, W, and star states using the 
R\'enyi  relative entropy of entanglement is given in Sec. \ref{MonogamyStates}.  The tripartite thermal entanglement of various models is 
characterized in Sec. \ref{XYZ} and \ref{planar}.  Then, the distribution of entanglement for the spin models is discussed in 
Sec. \ref{Distribution}. Discussion of the results and concluding remarks are given in Sec. \ref {Conclusions}.

\section{Results}
\subsection{Entanglement measures in terms of R\'enyi relative entropy}\label{measures}
The von Neumann entropy can be used to measure entanglement in bipartite systems \cite{nielsen2010quantum,byrnes2021quantum}.  The positive partial transpose (PPT) criterion 
\cite{HORODECKI1997333}, which gives rise to negativity, is only a sufficient condition when we look beyond two-qubit systems, giving a 
satisfactory measure only for $2 \times 2$ and $2 \times 3$ systems. Concurrence is an entanglement monotone that can be used to measure entanglement in 
pure and mixed two-qubit systems \cite{concur1}. As such, measures of entanglement such as von Neumann entropy, negativity, and 
concurrence can be 
satisfactorily used only to study bipartite systems.  

Measuring entanglement in tripartite and higher systems is generally a more difficult task.  For example, 
$n$-tangle \cite{ntangle} can measure genuine multipartite entanglement but cannot be used for tripartite systems with bipartite-like entanglement.  Additionally, measuring 
the tangle of mixed states is computationally hard due to the requirement of convex optimization.  Generalized concurrence is 
neither a sufficient nor necessary condition to compute entanglement \cite{Gilad}. 

An alternative approach is to use the geometric nature of the space of quantum systems to find the entanglement in the system.  In this approach, 
we use a geometric measure to find the distance between a given density matrix and the closest separable state.  This distance quantifies the amount
of entanglement in the system.  Here, a separable state is a convex combination of product state and has the following 
generic form: 
\begin{equation}
    \rho_{AB} = \sum_{i} p_{i} \, \rho_{A}^{i} \otimes \rho_{B}^{i} \,,
\label{separablestate}    
\end{equation}

\vspace{-2mm} 
\noindent where $p_{i}$ is the probability distribution and $\rho_{A}^{i}$  and  $\rho_{B}^{i}$ belong to the convex set of quantum states in 
$\mathcal{H}_{A}$ and $\mathcal{H}_{B}$, respectively. 

A standard distance measure that is used is based on the relative entropy.  The relative entropy between two classical probability distributions, $P = \{p_{1}, p_{2}, ... , p_{N} \}$  and 
$Q = \{q_{1}, q_{2}, ... , q_{N} \}$, is 
\begin{equation}
    S(P\|Q) = \sum_{i} p_{i} ( \ln p_{i} - \ln q_{i} ) \equiv - \sum_{i} p_{i} \ln q_{i} - S(P),
\label{classicalRE}    
\end{equation}

\vspace{-2mm} 
\noindent
where $S(P) =  - \sum_{i} p_{i} \ln p_{i}$ is the Shannon entropy. The quantum generalization of relative entropy based on von Neumann entropy
$S(\rho) = - \hbox{Tr} \rho \ln \rho$ is  
\begin{equation}
    S(\rho \| \sigma) = \hbox{Tr} \; \rho ( \ln \rho - \ln \sigma) \equiv - \hbox{Tr} \; \sigma \ln \rho - S(\rho)\,,
\label{quantumRE}    
\end{equation}
where $\rho$ and $\sigma$ are two density matrices.  The relative entropy of entanglement finds the distance to the closest separable 
state and is defined as
\begin{equation}
    E(\rho) \equiv \min_{\sigma \in \mathcal{D}}S(\rho \| \sigma) = \min_{\sigma \in \mathcal{D}} 
    \hbox{Tr} \; \rho \ln \frac{\rho}{\sigma}\,,
\end{equation}
where $\rho$ is the density matrix with entanglement, $\sigma$ is the separable state, and $\mathcal{D}$ is the set of all separable states.  
The minimization of the expression determines the closest separable state.  

The relative entropy of entanglement typically uses the von Neumann entropy in its definition. However, the von Neumann entropy is useful only in situations 
where the law of large numbers applies.  In non-ergodic and non-asymptotic settings \cite{renyi1961measures}, the law of large numbers does not hold.  In such situations, one uses other entropy measures like the min-entropy, the max-entropy, and the collision 
entropy \cite{muller2013quantum} given respectively by 
\begin{eqnarray}
    S_{\rm min} &=&  - \ln \| \rho \| \,,
    \label{minentropy} \\
    S_{\rm max} &=&  \ln {\hbox{rank}} (\rho) \,,
    \label{maxentropy} \\
    S_{\hbox{c}} &=& - \ln \hbox{Tr} [ \rho^{2} ]\,.
    \label{collisionentropy} 
\end{eqnarray}
Here $\| . \|$ refers to the operator norm.  The von Neumann entropy and the set of entropies in Eqs. \eqref{minentropy} - \eqref{collisionentropy} 
can be unified using a parametrized entropy known as the R\'enyi entropy
\begin{equation}
    S_{\alpha}(\rho)  =  \frac{1}{1 - \alpha} \ln \hbox{Tr} [ \rho^{\alpha} ]\,.
    \label{Renyientropy}
\end{equation}
Here $\alpha \in (0,1) \cup (1,\infty)$ is the generalization parameter which allows to unify all these entropies \cite{traditional1,traditional2}.  
From the expression of the R\'enyi entropy, the min-max entropies, the collision entropy, and the von Neumann entropies can be recovered via the limits
\begin{eqnarray}
    \lim_{\alpha \rightarrow \infty}  S_{\alpha}(\rho)  &=& S_{\rm min} (\rho) ,
    \label{limitminentropy} \\
    \lim_{\alpha \rightarrow 0}  S_{\alpha}(\rho)  &=& S_{\rm max} (\rho) ,
    \label{limitmaxentropy} \\
    \lim_{\alpha \rightarrow 2}  S_{\alpha}(\rho)  &=& S_{c} (\rho)  ,
    \label{limitcollisionentropy} \\
    \lim_{\alpha \rightarrow 1}   S_{\alpha}(\rho)  &=& S(\rho)  .
    \label{limitvonNeumannentropy}
\end{eqnarray}
Using the R\'enyi entropy in Eq. (\ref{Renyientropy}), one can generalize the relative entropy to give the corresponding expression 
\begin{equation}
    S_{\alpha}^{T}(\rho \| \sigma) \equiv \frac{1}{1-\alpha}  \ln [ {\rm Tr} ( \rho^{\alpha} \sigma^{1-\alpha} ) ]\, ,
    \label{traditionalRenyi}
\end{equation}
where the range of the R\'enyi parameter is $\alpha \in (0,1) \cup (1,\infty)$.  One of the important properties of an entanglement 
measure is that it should be convex.  In the case of a distance-based entanglement measure, the function should be jointly convex, i.e.,
convex with respect to the density matrix under consideration and the reference state.   For the R\'enyi relative entropy of entanglement, 
the joint convexity holds only for the range $0 \leq \alpha \leq 2$. 

An alternative definition of R\'enyi relative entropy known as 
sandwiched R\'enyi relative entropy was introduced in Ref. \cite{inequality2,inequality4} to increase the range of validity of the 
generalization parameter $\alpha$. The expression corresponding to the sandwiched R\'enyi relative entropy is 
\begin{equation}
    S_{\alpha}^{S} \equiv \frac{1}{1-\alpha}  
    \ln  \big[ {\rm Tr} \big(\sigma^{\frac{1-\alpha}{2 \alpha}} \rho \sigma^{\frac{1-\alpha}{2 \alpha}}\big)^{\!\alpha} \big]\,.     
    \label{SandwichedRenyi}
\end{equation}
The joint convexity requirement enforces the range on $\alpha$ to be $1/2 \leq \alpha \leq \infty$.
We use Eq. \eqref{traditionalRenyi} and Eq. \eqref{SandwichedRenyi} to investigate the entanglement and its distribution in three-qubit spin systems.  
In the parametric regime $0 \leq \alpha \leq 1/2$, only the traditional R\'enyi relative entropy is useful, and in the region
$2 \leq \alpha \leq \infty$ only the sandwiched R\'enyi relative entropy is useful.  In the region $1/2 \leq \alpha \leq 2$,
both the traditional and sandwiched R\'enyi relative entropies can be used for computing the entanglement in the system.

\begin{figure}[t] 
\includegraphics[width=0.95\columnwidth]{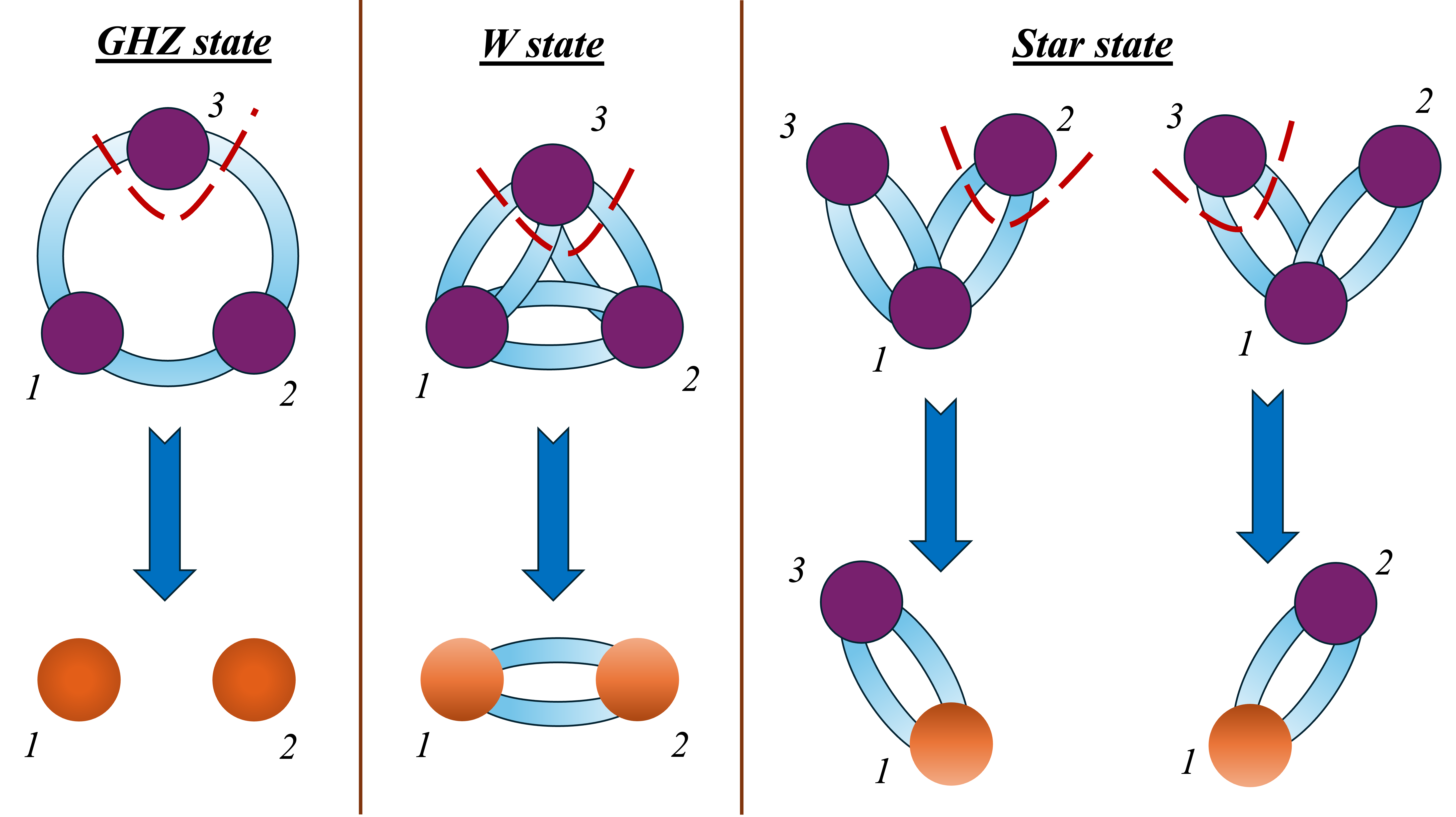}
\caption{A schematic illustration of tripartite entangled states is given above for GHZ state, W state, and star state. The qubits are denoted 
by circles and the entanglement is represented by blue-colored closed bands running between the qubits.  In the GHZ state, there are three qubits
lying within the closed band, indicating that the qubits are entangled in a tripartite way.  There are only two qubits in a closed band in the 
W state and the star state which implies that the entanglement in these states are bipartite.  The dashed red line around a qubit attributes 
a tracing out operation over it, and the resulting bipartite state is given below. The purple-colored qubits are the state
in the original tripartite pure state, and the orange-colored qubits are those that have lost entanglement due to tracing out operation. }
\label{tripartitestatesShematic}
\end{figure}

\begin{figure}[t] 
\centering
\includegraphics[width=0.95\columnwidth]{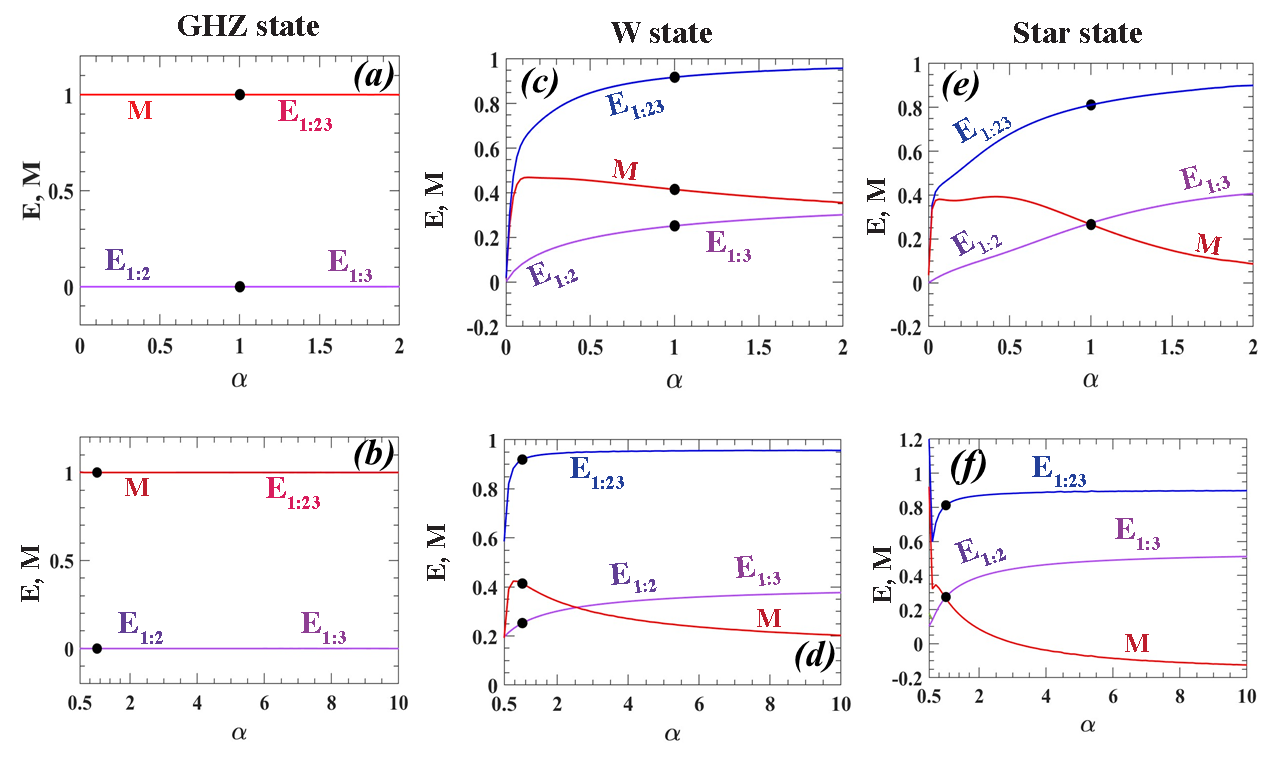}
\caption{Entanglement and monogamy versus $ \alpha $ computed using the traditional and sandwiched R\'enyi relatives entropy for the 
GHZ state in plots (a) and (b), for the W states in (c) and (d) and the star states in (e) and (f).   The plots shows the variation of 
the entanglement $E_{1:23}$, the bipartite entanglements $E_{1:2}$ and $E_{1:3}$ and the monogamy of entanglement $M$.  The values of the 
bipartite entanglements $E_{1:2}$ and $E_{1:3}$ are equal for all three states, 
and hence they completely overlap. The monogamy of entanglement in the $\alpha \rightarrow 1$ limit, i.e., corresponding to the 
von Neumann entropy is indicated by a dot for the sake of comparison. 
}
\label{RenyiStates}
\end{figure}

\subsection{Monogamy of entanglement in tripartite states}\label{MonogamyStates}

It is widely considered that the entanglement in tripartite systems can be classified into two types \cite{dur2000three}, namely the {\it(i)} GHZ class and the {\it (ii)} W class, depending on Local Operations and Classical Communication (LOCC).  In the GHZ class, the states are entangled in such a way that the 
loss of even a single qubit in the tripartite state leads to a totally disentangled state.  Meanwhile, in the W class, the decoherence of 
one qubit leaves behind a mixed entangled state.  This observation leads us to conclude that the entanglement in a GHZ state is distributed in a 
tripartite fashion, whereas in the W state, it is in a bipartite manner.   

The entanglement distribution can also be understood using the monogamy relations, first introduced in Ref. \cite{monogamy2} for tripartite systems and later generalized for multipartite systems in Ref.\cite{Osbornemultimono}. 
The monogamy relation for a tripartite system is 
\cite{monogamy2,monogamy3}
\begin{equation}
E(\rho_{1:23}) \geq E(\rho_{1:2})+E(\rho_{1:3})\,,
\end{equation}
where $E(\rho_{1:23})$ is the bipartite entanglement between the qubit $1$ and the block $23$, while $E(\rho_{1:2})$ and $E(\rho_{1:3})$ 
are, respectively, the bipartite entanglement in the reduced systems $\rho_{12}$ and $\rho_{13}$.  If this inequality is obeyed, the 
system is said to be monogamous; otherwise, the system is said to be polygamous. The monogamy relation holds for all entanglement 
measures, but in our work we use the R\'enyi relative entropy of entanglement.  
For the sake of analyzing a given tripartite state, we rewrite the monogamy of entanglement expression as 
\begin{equation}\label{MonogamyStatesRelation}
    M \equiv E(\rho_{1:23})-E(\rho_{1:2})-E(\rho_{1:3})\,.
\end{equation}
If $M$ is non-negative, then the distribution is monogamous; otherwise, the entanglement is distributed polygamously. 

Evaluating the monogamy of entanglement for some prototypical states using the R\'enyi relative entropy entanglement measure already 
yields some interesting results, which we now discuss.  
First we 
look into the two symmetric states, namely the GHZ and W states 
\begin{align}
    |\text{GHZ} \rangle  &  =  \tfrac{1}{\sqrt{2}} ( |000 \rangle + |111 \rangle ),  \nonumber \\
    |W \rangle  & =  \tfrac{1}{\sqrt{3}} (|001 \rangle + |010 \rangle + |100 \rangle) .
\label{ghzwstates}    
\end{align}

Let us begin by considering the case of GHZ states. From the plots in Fig. \ref{RenyiStates} (a) and (b), we find that the monogamy of entanglement remains 
positive for all valid values of $\alpha$ for both the regular R\'enyi relative entropy and the sandwiched R\'enyi relative entropy.  
This happens because $E_{1:23}$ is positive since it measures the entanglement between the system $1$ and the bipartite block $23$.  
We observe from the plot Fig \ref{RenyiStates} (a) and (b), that the entanglements $E_{1:2}$ and $E_{1:3}$ are both zero, 
since the bipartite reduced state $(|00 \rangle \langle 00| + |11 \rangle \langle 11|)/2$ is a separable state.  A schematic illustration 
is given in Fig. \ref{tripartitestatesShematic}, showing that tracing out of a qubit destroys all the entanglement in the system.

Next, we focus on the $W$ state and find that both $E_{1:23}$ and the bipartite entanglements 
have positive values (see Fig. \ref{RenyiStates} (c) and (d)). This is as expected as $W$ states are well known to retain entanglement when tracing out one of the qubits.  
Explicitly, we may see this since
\begin{equation}
    \rho_{12}^{W} = {\rm Tr}_{3} ( \rho^{W}_{123} ) = \rho_{13}^{W} = {\rm Tr}_{2}( \rho^{W}_{123} ) = \tfrac{1}{3} | 00 \rangle \langle 00 | + \tfrac{2}{3} | \Phi^+ \rangle \langle \Phi^+ |  \label{reducedwstates}
\end{equation}
where $ | \Phi^+ \rangle = \tfrac{1}{\sqrt{2}}(|00 \rangle + |11 \rangle)$.  The bipartite state is thus an entangled mixed state. 
Through the schematic illustration in Fig. \ref{tripartitestatesShematic}, we show the entanglement distribution in the tripartite W state and 
the entanglement present after the tracing out of one of the qubits.  Surprisingly, the monogamy of entanglement of this state is positive, 
as seen in Fig. \ref{RenyiStates} (c) and (d) since the 
entanglement $E_{1:23}$ is greater than the combined value of $E_{1:2}$ and $E_{1:3}$. 

As a contrasting case, it is interesting to study the monogamy of entanglement of the star state \cite{Star1,Star2,radhakrishnan2019dynamics} 
\begin{equation*}
    |\rm{star} \rangle = \tfrac{1}{2} (|000 \rangle + |001 \rangle + |101 \rangle + |111 \rangle).
\end{equation*}
In a three-qubit star state, the qubits can be classified into two types, namely the {\it(i)} the central qubit which is entangled to the other two 
qubits and {\it(ii)} the peripheral qubit which is entangled to only the central qubit. Here, we take qubit 1 as the central qubit and qubits 2 and 3 as 
the peripheral qubit.  The form of the star state is shown in 
Fig. \ref{tripartitestatesShematic}, where we see entanglement in the qubit pairs $12$ and $13$, but the qubit pair $23$ is not entangled.  
The reduced density matrices of the states  $\rho_{12}$ and $\rho_{13}$ are 
\begin{align}
\rho_{12}^{s} & = {\rm Tr}_{3} ( \rho^{S}_{123} ) = \tfrac{1}{4} (2 + \sqrt{2})  | s_{12}^+ \rangle \langle s_{12}^+ | 
                   + \tfrac{1}{4} (2 - \sqrt{2})  | s_{12}^- \rangle \langle s_{12}^- | \,,
                   \label{rhostar}
\end{align}
where
\begin{align}
| s_{12}^\pm \rangle = \tfrac{1}{\sqrt{2}} (| 00  \rangle \pm | +1 \rangle )  
\end{align}
are partially entangled states.  For subsystem 13, we have the state 
\begin{align}
\rho_{13}^{s} & = {\rm Tr}_{2} ( \rho^{S}_{123} ) = \tfrac{1}{4} (2 + \sqrt{2})  | s_{13}^+ \rangle \langle s_{13}^+ | 
                   + \tfrac{1}{4} (2 - \sqrt{2})  | s_{13}^- \rangle \langle s_{13}^- | \label{rhostar13}
\end{align}
where
\begin{align}
| s_{13}^\pm \rangle = \tfrac{1}{\sqrt{2}} (| 0+  \rangle \pm | 11 \rangle )  .
\end{align}

The variation of monogamy for the star state with respect to the R\'enyi generalization parameter $\alpha$ is shown in Fig. \ref{RenyiStates}(e) for the 
regular R\'enyi relative entropy and in Fig. \ref{RenyiStates}(f) for the sandwiched R\'enyi relative entropy.  In Fig. \ref{RenyiStates}(e) we see that the monogamy of entanglement is always positive for the regular R\'enyi relative entropy 
of entanglement.  But the regular R\'enyi relative entropy is jointly convex only in the range $0 \leq \alpha \leq 2$.  To 
check for further values of the generalization parameter $\alpha$ we compute the entanglement using the sandwiched R\'enyi relative entropy
since this measure is jointly convex in the range  $0.5 \leq \alpha \leq \infty$.  The results corresponding to the 
sandwiched R\'enyi relative entropy based entanglement measure are shown in Fig. \ref{RenyiStates} (f), where we observe a polygamous
nature for values of $\alpha > 2$.   

The reason the star state is more polygamous than the $ W $ state according to the R\'enyi relative entropy can be understood as follows. The entanglement in a $W$ state is distributed in a bipartite fashion where the entanglement exists between the qubit pairs 
$12$, $13$ and $23$.  When the qubit $3$ is traced out, two entanglement bonds between qubits $13$ and $23$ are lost and, 
consequently, there is a high level of mixedness in the reduced state $\rho_{12}^{W}$. 
Comparing (\ref{rhostar}) and (\ref{reducedwstates}), we see that when tracing out qubit 2, the purity of the star state is higher,
i.e., $ \tfrac{1}{4} (2 + \sqrt{2}) \approx 0.854 > 2/3 $.  We may understand this heuristically in Fig. \ref{tripartitestatesShematic} as arising from the fact that only one entanglement bond is broken in a star state, whereas in a $W$ state two bonds must be broken.  
Hence while the entanglement in the bipartite block $E_{1:23}$ is higher than the 
reduced state entanglement $E_{1:2}$ and $E_{1:3}$, for certain values of the R\'enyi generalization parameter $\alpha$ 
we observe the condition $E_{1:23} < E_{1:2} + E_{1:3}$ resulting in negative values for the monogamy of entanglement for the star state.  
The value of $M$ tends to be more negative for larger $ \alpha $ since it is well known that the R\'enyi parameter biases the probabilities in an entropy. 
Thus in an asymmetrically entangled tripartite state, we find that the monogamy of entanglement can take negative values when we use a 
R\'enyi relative entropy based entanglement measure.

\begin{figure}[h] 
\includegraphics[width=0.90\columnwidth]{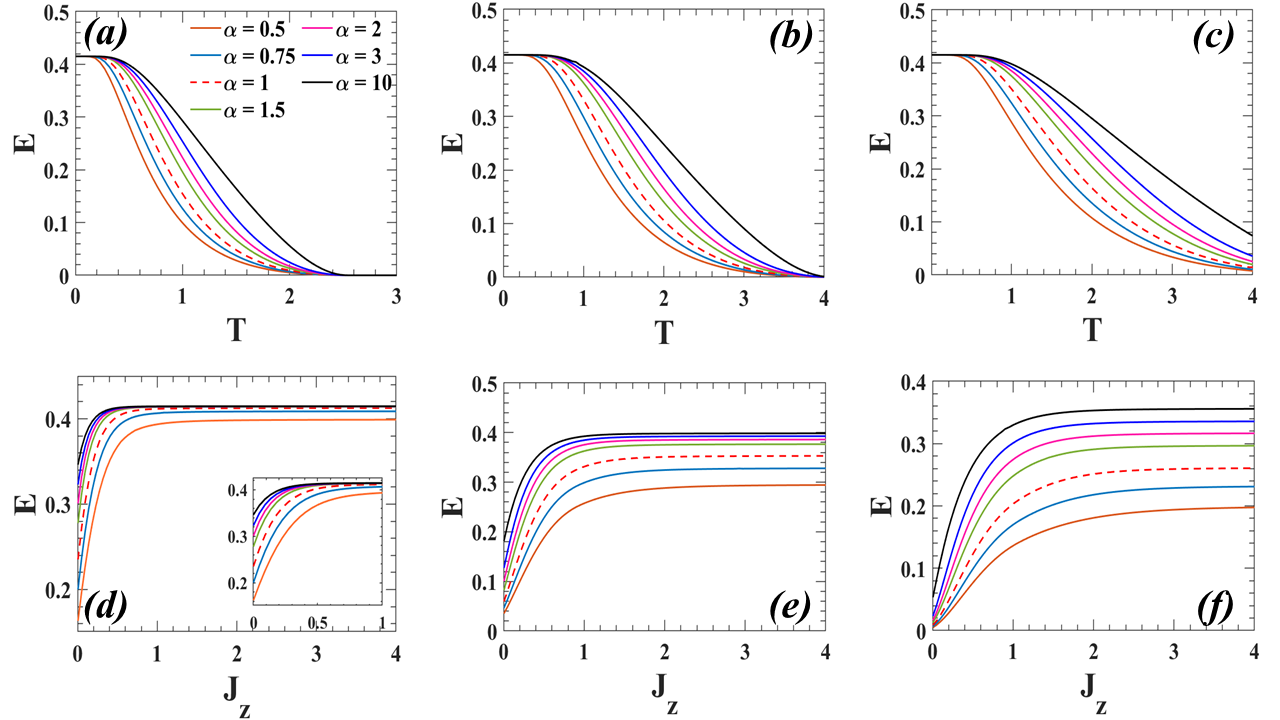}
\caption{The tripartite entanglement of the XYZ model measured using sandwiched R\'enyi relative entropy for different values of $\alpha$
is given for the spin-spin interaction parameters (a) $J_{z} = 0.5$, (b) $J_{z} = 1.0$ and (c) $J_{z} = 1.5$.  The variation of the entanglement 
with respect to the temperature $T$ is given for (d) $T = 0.5$, (e) $T = 1$ and (f) $T = 1.5$.   The inset of (d) shows an enlarged region in the limit of small $J_z$.
The values of the other parameters are $J_x = 0.8$ and $J_y = 0.5$.  The dashed line corresponds to the von Neumann entropy-based relative entropy of 
entanglement $\left(\alpha=1 \right)$.}
\label{SandwichedXYZ}
\end{figure}

\subsection{Entanglement in Heisenberg XYZ and XXZ models}\label{XYZ}

We now examine the entanglement in some physical models, specifically the three-site Heisenberg XYZ and XXZ models. The two site models have been well studied through the works in 
\cite{BipartiteXYZ,BipartiteXYZ1,BipartiteXYZ2,ThermalXXZ}.  In models with more than two sites, entanglement has been calculated only 
for the reduced bipartite systems.  In our work we find the total entanglement in the system using the R\'enyi relative entropy of entanglement. 
To cover the entire range of the parameter $\alpha$ we use both the traditional and sandwiched version of the R\'enyi relative entropy. The 
entanglement is characterized by varying several control parameters like temperature, spin-spin coupling strength, and the anisotropy parameter.

\subsubsection{The XYZ model}
We consider a Heisenberg chain of three spin- $1/2$ particles labeled with site index $i$ and interacting only with its nearest neighbors. The corresponding Hamiltonian is given by 
\begin{eqnarray}
    H=\sum_{i=1}^{3}J_x \sigma_i^x \sigma_{i+1}^x + J_y \sigma_i^y \sigma_{i+1}^y + J_z\sigma_i^z \sigma_{i+1}^z\,,
    \label{HXYZ}
\end{eqnarray}
where $J_l > 0 \left(l=x,y,z\right)$ is the anti-ferromagnetic exchange coupling along the $l^{th}$ direction and $\left\{\sigma_i^x, \sigma_i^y, 
\sigma_i^z\right\}$ are the Pauli matrices for the $i^{th}$ qubit. By adjusting the coupling constants $J_l$, one may alter this model to 
describe the anisotropic XYZ model when $J_x \neq J_y \neq J_z $, XXZ model with $J_x = J_y \neq J_z$ and the isotropic XXX model with $J_x = J_y = J_z$.  In 
all of the models considered in this paper, we assume periodic boundary conditions.  Details of the calculation to obtain the thermal density matrix for the Hamiltonian is given in the Methods.

\begin{figure}[t] 
\includegraphics[width=0.90\columnwidth]{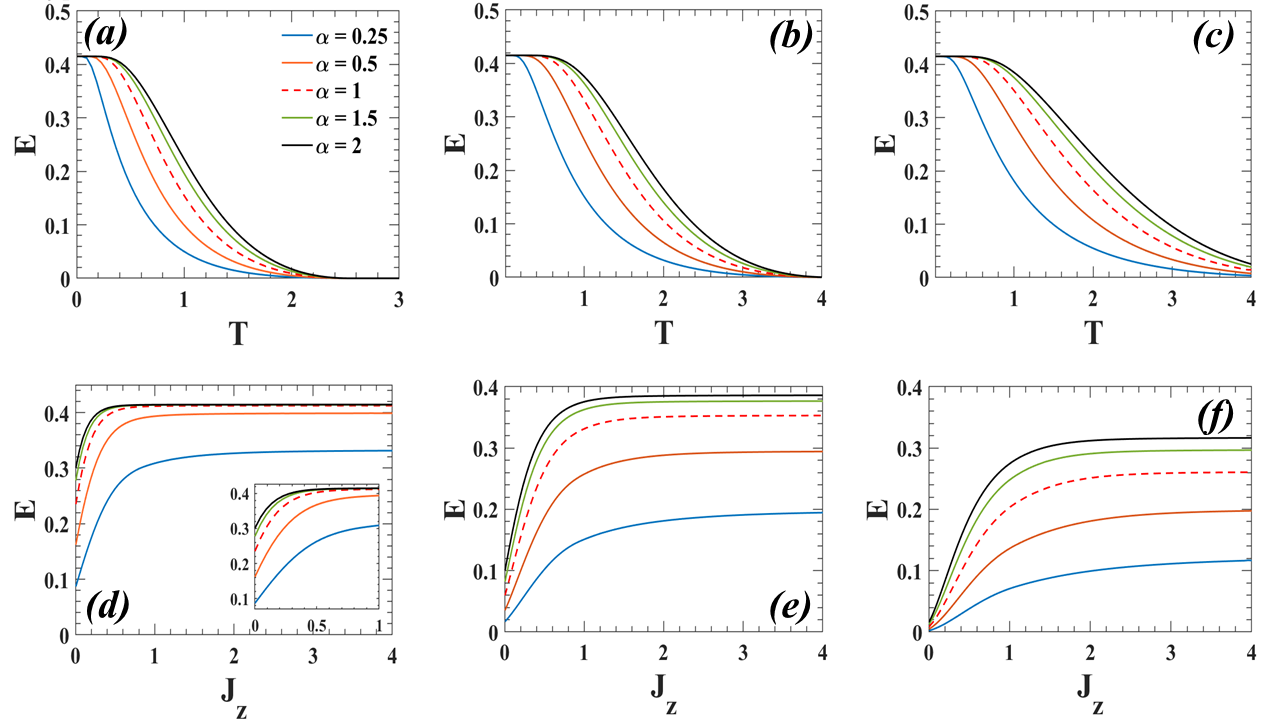}
\caption{The tripartite entanglement of the XYZ model measured using traditional R\'enyi relative entropy for different values of $\alpha$
is given for the spin-spin interaction parameters (a) $J_{z} = 0.5$, (b) $J_{z} = 1.0$ and (c) $J_{z} = 1.5$.  The variation of the entanglement 
with respect to the temperature $T$ is given for (d) $T = 0.5$, (e) $T = 1$ and (f) $T = 1.5$.   The inset of (d) shows an enlarged region in the limit of small $J_z$.
The values of the other parameters are $J_x = 0.8$ and $J_y = 0.5$.  The dashed line corresponds to the von Neumann entropy-based relative entropy of 
entanglement $\left(\alpha=1 \right)$.}
\label{RenyiXYZ}
\end{figure}

The tripartite entanglement of the model is calculated using traditional R\'enyi relative entropy, and the results are shown in Fig. \ref{RenyiXYZ}. 
In the plots Fig. \ref{RenyiXYZ} (a-c), the variation of the entanglement with $T$ is shown for fixed values of $J_z$.  The entanglement remains 
almost constant in the low-temperature limit, and beyond a particular temperature it decreases and finally becomes zero at a particular temperature
$E(T) = 0$,  $\forall T \geq T_{c}$ where $T_{c}$ is the critical temperature. 
The fall in entanglement with temperature is due to the additional decoherence that is present for higher temperatures.  Here we note that 
with an increase in the R\'enyi parameter $\alpha$, the entanglement falls slower.  Thus we can conclude that the R\'enyi parameter gives a larger numerical 
value for a given amount of entanglement for $\alpha > 1$. Further, we also find that when the value of the coupling parameter is increased, the critical
temperature also rises.   
The variation of the entanglement with the coupling strength is shown in Fig. \ref{RenyiXYZ} (d-f) for different values 
of temperature $T$ and the $\alpha$.  The system's entanglement slowly increases against $J_{z}$ and reaches a saturation value.  The saturation value
and the rate of increase of entanglement are proportional to the R\'enyi parameter $\alpha$.  For the R\'enyi relative entropy of entanglement, 
we observe a higher saturation value of entanglement for $\alpha > 1$ in comparison with the relative entropy based on von Neumann entropy ($\alpha = 1$).  
In Fig. \ref{SandwichedXYZ} (a-f), we do the same study for the sandwiched R\'enyi relative entropy whose range of the generalization parameter is 
$\alpha \in [\frac{1}{2}, \infty)$. These results agree with the corresponding ones obtained using the traditional R\'enyi relative entropy,
and the effects of slower decay of entanglement and higher saturation value are confirmed for $\alpha > 2$.

\subsubsection{The XXZ model}

The XXZ model can be obtained from Eq. (\ref{HXYZ}) by setting the spin-spin interaction to $J_x=J_y=J$ and $J_z/J=\Delta$. 
Explicitly, the Hamiltonian is
\begin{eqnarray}
    H=\sum_{i=1}^{3}J \left(\sigma_i^x \sigma_{i+1}^x + \sigma_i^y \sigma_{i+1}^y + \Delta \sigma_i^z \sigma_{i+1}^z\right),
    \label{HXXZ}
\end{eqnarray}
where $\Delta$ is the anisotropy of the spin–spin interaction in the $z$ direction. This model is exactly solvable using the Bethe ansatz and 
possesses three distinct ground state quantum phases: a ferromagnetic Ising phase for $\Delta \leqslant -1$, a gapless Luttinger-liquid phase for $-1<\Delta 
\leqslant 1$ and a gapped N\'eel phase exhibiting long-range order for $\Delta > 1$. It is known that the ferromagnetic phase spontaneously breaks $\mathbb{Z}_2$ 
spin-reflection symmetry \cite{phasesXXZ,phasesXXZ1,phasesXXZ2,phasesXXZ3}. Consequently, the fully polarized ground state is separable with vanishing 
entanglement. Hence we will restrict our attention to the case of $\Delta>-1$.  Details of the calculation to obtain the thermal density matrix for 
the Hamiltonian are given in the Methods.  

\begin{figure}[hpbt] 
\includegraphics[width=0.90\columnwidth]{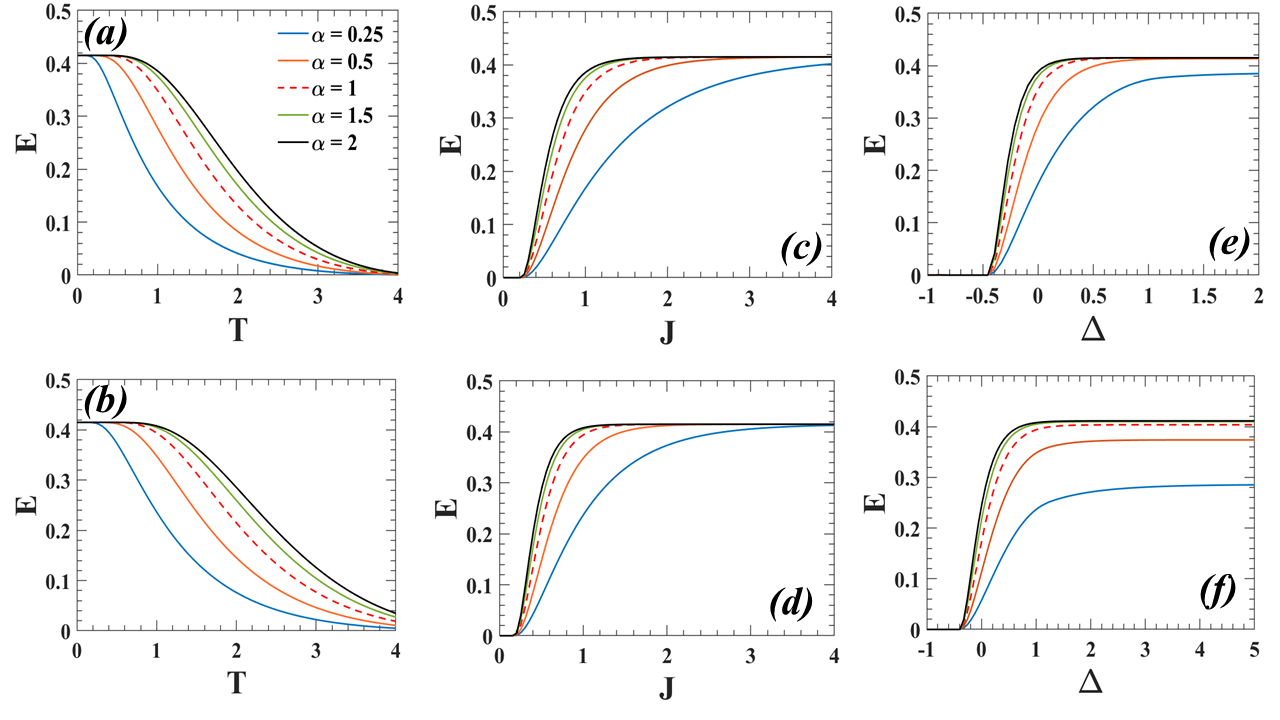}
\caption{The traditional Renyi relative entropy of entanglement of the XXZ Heisenberg model as a function of the temperature  
is shown for (a) $\Delta = 0.5$ and (b) $\Delta = 1.0$ for $J = 1$.  The variation of entanglement as a function of the interaction parameter $J$
is given for (c) $\Delta = 0.5$ and (d) $\Delta = 1.0$ with the temperature being $T = 1$. Finally we show the variation of the anisotropy parameter
$\Delta$ for (e) $T = 0.5$ and (f) $T = 1$ with the interaction parameter being set at $J = 1$.}
\label{RenyiXXZ}
\end{figure}

\begin{figure}[hpbt] 
\includegraphics[width=0.90\columnwidth]{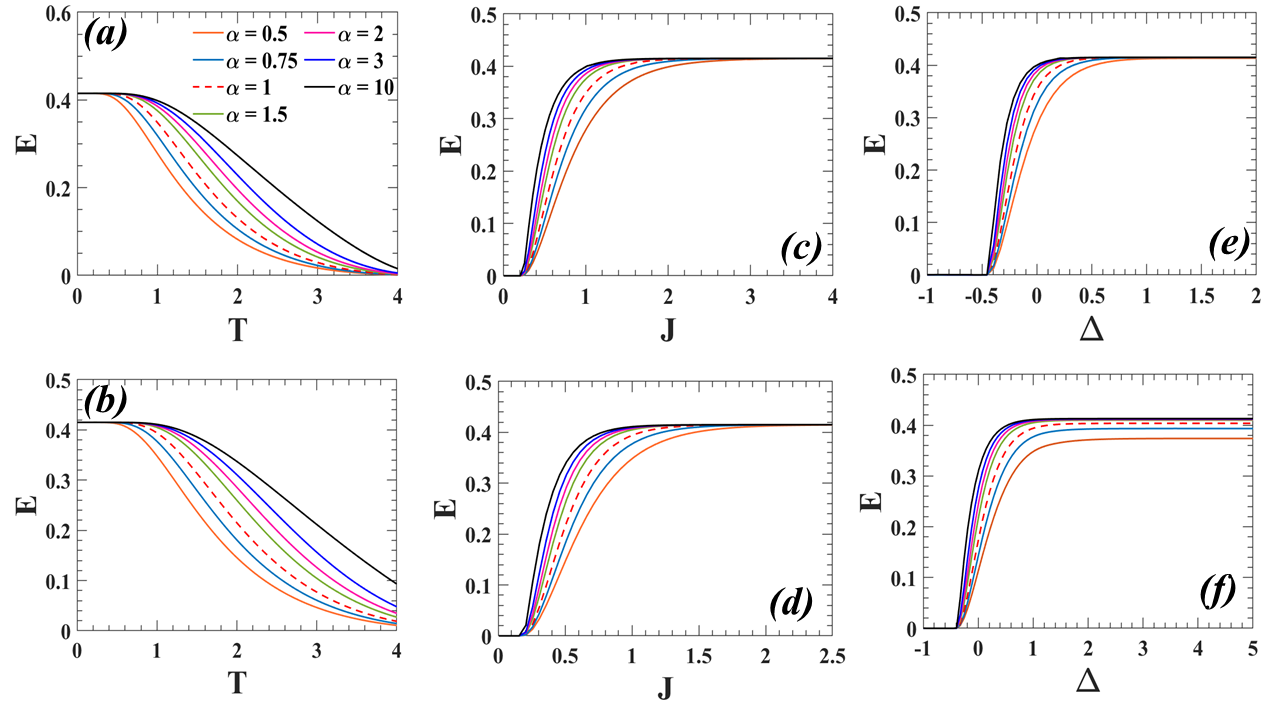}
\caption{The sandwiched Renyi relative entropy of entanglement of the XXZ Heisenberg model as a function of the temperature  
is shown for (a) $\Delta = 0.5$ and (b) $\Delta = 1.0$ for $J = 1$.  The variation of entanglement as a function of the interaction parameter $J$
is given for (c) $\Delta = 0.5$ and (d) $\Delta = 1.0$ with the temperature being $T = 1$. Finally we show the variation of the anisotropy parameter
$\Delta$ for (e) $T = 0.5$ and (f) $T = 1$ with the interaction parameter being set at $J = 1$.}
\label{SandwichedXXZ}
\end{figure}

The entanglement of the three site XXZ model as a function of temperature is shown in Fig. \ref{RenyiXXZ} (a) and (b) for fixed 
values of the anisotropy parameter $\Delta$ and the R\'enyi parameter $\alpha$.  The plots show that the entanglement decreases as expected 
due to thermal decoherence.  Further, the critical temperature $T_c$ 
is significantly enhanced by increasing of $\Delta$ and $\alpha$.  
The entanglement with the spin-spin coupling parameter $J$ is shown in Fig. \ref{RenyiXXZ} (c) and (d).  The tripartite entanglement increases from
zero and attains a saturation value very rapidly.  The rate of increase of the entanglement differs based on the value of $\alpha$, but the finite 
saturation value attained is the same for all $\alpha$. The result obtained has been verified for higher values of $\alpha$ using the sandwiched 
R\'enyi relative entropy in Fig. \ref{SandwichedXXZ} (c) and (d).  The dependence of the entanglement with respect to the anisotropy parameter $\Delta$ is
shown for the traditional R\'enyi entropy and the sandwiched R\'enyi entropy in Fig. \ref{RenyiXXZ} (e) and (f) and \ref{SandwichedXXZ} (e) and (f), 
respectively. Here, we find that the entanglement increases and saturates. 

In the region  $-1<\Delta \leq 1$,  the entanglement increases monotonously with the anisotropy parameter.  This observation is similar to the results reported in 
Ref. \cite{ThermalXXZ} where the authors have investigated the thermal entanglement of the XXZ model.  In that work, they use concurrence to measure the 
pairwise entanglement of a three-qubit system and find that the entanglement drops to zero for $\Delta \gg 1$.  This is in contrast to our results, 
where we find entanglement for large values of $\Delta$.  This discrepancy occurs because the concurrence used in Ref. \cite{ThermalXXZ} detects the 
bipartite entanglement and fails to detect the tripartite entanglement in the system.  In the large $\Delta$ limit, the ground state of the system 
is the N\'eel state which is a three-qubit GHZ state. The entanglement in such a state can be detected and characterized only by a multipartite 
entanglement measure. This observation clearly underlines the need to use a multipartite entanglement measure, since a bipartite measure might 
not detect all the entanglement in the system.


\subsection{Entanglement in planar spin models}\label{planar}
\subsubsection{The XY model}

We now turn our attention to the three-site XY model with periodic boundary conditions. The Hamiltonian of the model is 
\begin{eqnarray}
    H=J\sum_{i=1}^{3} \left[\frac{1+\gamma}{2} \sigma_i^x \sigma_{i+1}^x + \frac{1-\gamma}{2} \sigma_i^y \sigma_{i+1}^y \right].
    \label{HXY}
\end{eqnarray}
The range of the anisotropy parameter $\gamma$ is restricted to the regime $[0,1]$ with the limiting cases $\gamma=0$ is the XX model and 
$\gamma=1$ is the Ising model. The XY model is one of the solvable models where its bipartite entanglement \cite{XYentanglemnet,XYentanglemnet1} and coherence \cite{XYcoherence,XYcoherence2,XYcoherence3} has been widely 
studied through a number of different methods. From the density matrix, we can compute the finite temperature density matrix.  The entanglement
of the density matrix has been calculated in the range $\alpha \in [0, 2]$ using the traditional R\'enyi relative entropy of entanglement 
and in the regime $\alpha \in [1/2, \infty)$ with the sandwiched R\'enyi relative entropy of entanglement.

\begin{figure}[t] 
\includegraphics[width=0.90\columnwidth]{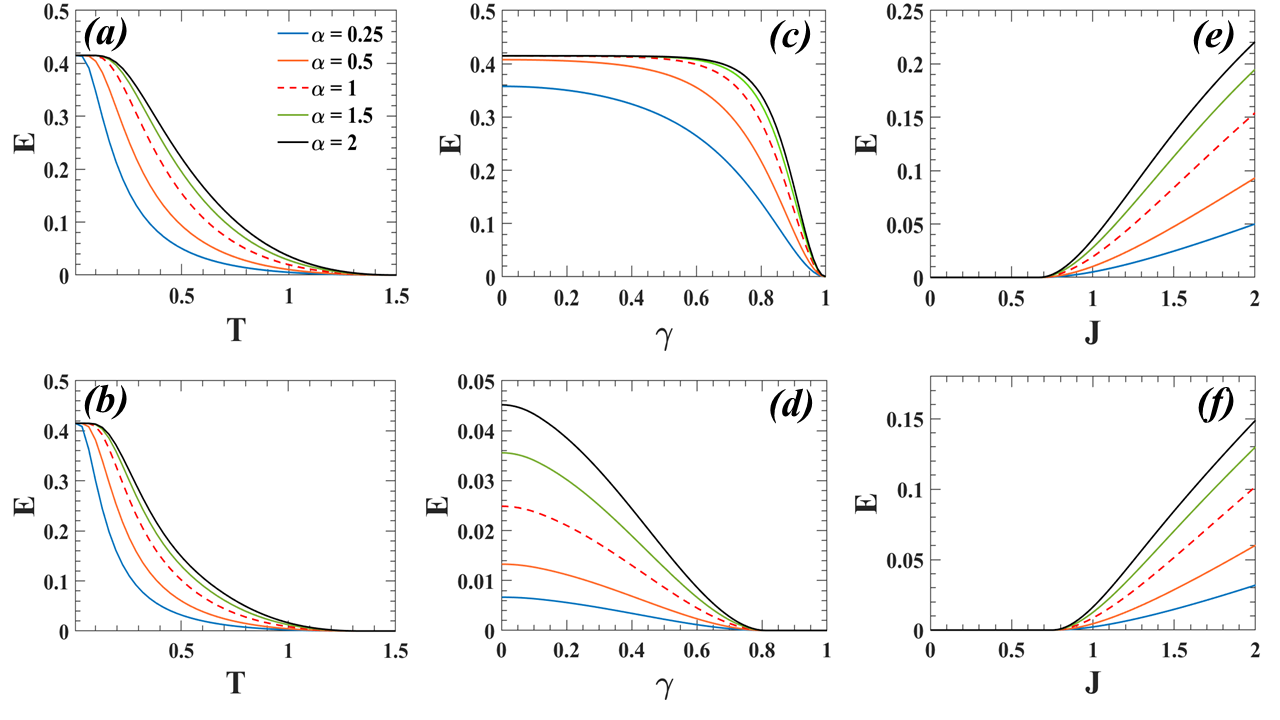}
\caption{The tripartite entanglement in the XY model obtained using the traditional R\'enyi relative entropy is characterized for different values of $
\alpha$.  The entanglement versus temperature $T$ for (a) $\gamma = 0.25$  and  (b) $\gamma = 0.5$ is given for $J=1$.  The variation of entanglement with $
\gamma$  for (c) $T = 0.1$ and (d) $T = 1$ is shown for $J=1$. The entanglement change with the interaction parameter $J$  (e) $\gamma = 0.25$ and (f) $
\gamma = 0.5$ is given in $T = 1$.}
\label{RenyiXY}
\end{figure}

\begin{figure}[h] 
\includegraphics[width=0.95\columnwidth]{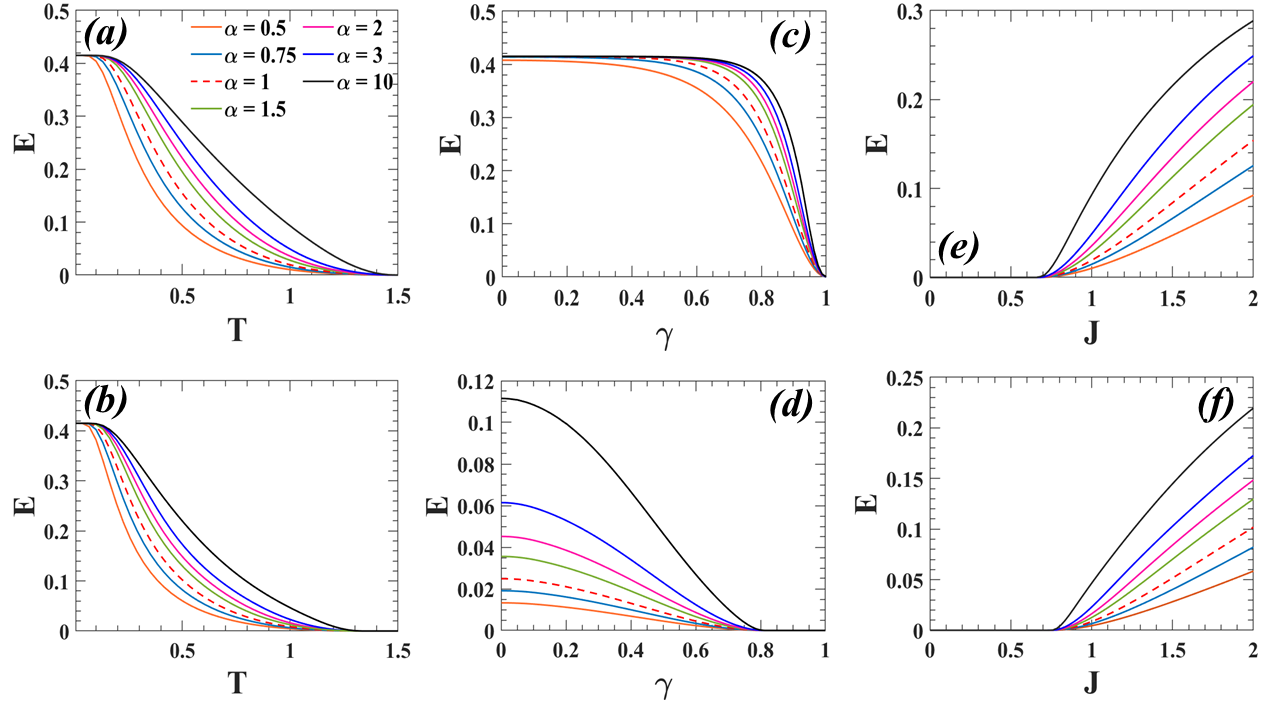}
\caption{The tripartite entanglement in the XY model obtained using the sandwiched R\'enyi relative entropy is characterized for different values of $
\alpha$.  The entanglement versus temperature $T$ for (a) $\gamma = 0.25$  and  (b) $\gamma = 0.5$ is given for $J=1$.  The variation of entanglement with $
\gamma$  for (c) $T = 0.1$ and (d) $T = 1$ is shown for $J=1$. The entanglement change with the interaction parameter $J$  (e) $\gamma = 0.25$ and (f) $
\gamma = 0.5$ is given in $T = 1$.}
\label{SandwichedXY}
\end{figure}

The plot of entanglement as a function of temperature $T$ is shown in Fig. \ref{RenyiXY} (a-b) and \ref{SandwichedXY} (a-b) for the traditional and 
the sandwiched R\'enyi relative entropy for different values of R\'enyi parameter $\alpha$ and the anisotropy parameter $\gamma$.  The entanglement
decreases monotonically with temperature due to thermal decoherence.  The variation of the tripartite entanglement with respect to 
$\gamma$ for different values of $\left(\alpha, T\right)$ is given in \ref{RenyiXY} (c-d) and \ref{SandwichedXY} (c-d) for the traditional and 
sandwiched relative entropies respectively.  From the results we observe that the entanglement has a maximal value at $\gamma =0$ which is the 
isotropic XX model and goes to zero at $\gamma = 1$ which is the Ising model.  Finally in Fig. \ref{RenyiXY} (e-f) and \ref{SandwichedXY} (e-f) we 
show the entanglement as a function of spin-spin interaction parameter $J$ and we find that the amount of entanglement increases with $J$.  


\subsubsection{The transverse-field Ising model}
The Hamiltonian of the three-site transverse-field Ising model with the nearest-neighbor coupling under the effect of an external magnetic field 
in the $z$-direction is 
\begin{eqnarray}
    H= \sum_{i=1}^{3} \left( \lambda \sigma_i^x \sigma_{i+1}^x + \sigma_i^z \right),
    \label{HIsing}
\end{eqnarray}
where $\lambda = \frac{J}{B}$ is a dimensionless parameter measuring the strength of the longitudinal spin-spin coupling with respect to the external 
transverse magnetic field $B$.  It is well known that such models undergo a quantum phase transition at zero temperature for the critical value 
$\lambda_c = 1$, which separates the quantum paramagnetic phase from the ferromagnetic phase.  Entanglement has been successfully used to detect such 
phase transitions \cite{osterloh2}. Details of the calculation to obtain the thermal density matrix for the Hamiltonian are given in the Methods.

\begin{figure}[t] 
\includegraphics[width=0.90\columnwidth]{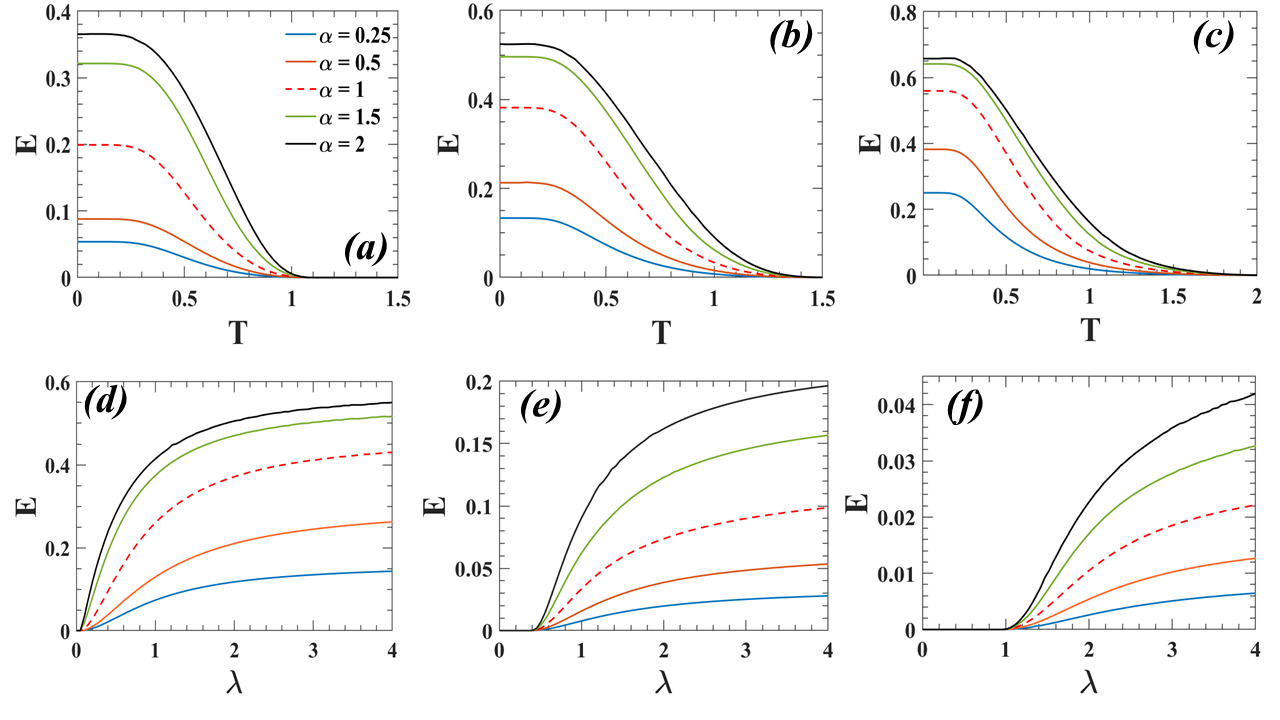}
\caption{Tripartite entanglement in the transverse-field Ising model calculated using the traditional R\'enyi relative entropy for 
various values of $\alpha$.  The entanglement variation with temperature for (a) $\lambda = 0.5$, (b) $\lambda = 1$ and (c) $\lambda = 2$.  
Entanglement versus $\lambda.  $ is plotted for (d) $T = 0.5$, (e) $T = 1$ and (f) $T = 1.5$.}
\label{RenyiIsing}
\end{figure}

\begin{figure}[t] 
\includegraphics[width=0.90\columnwidth]{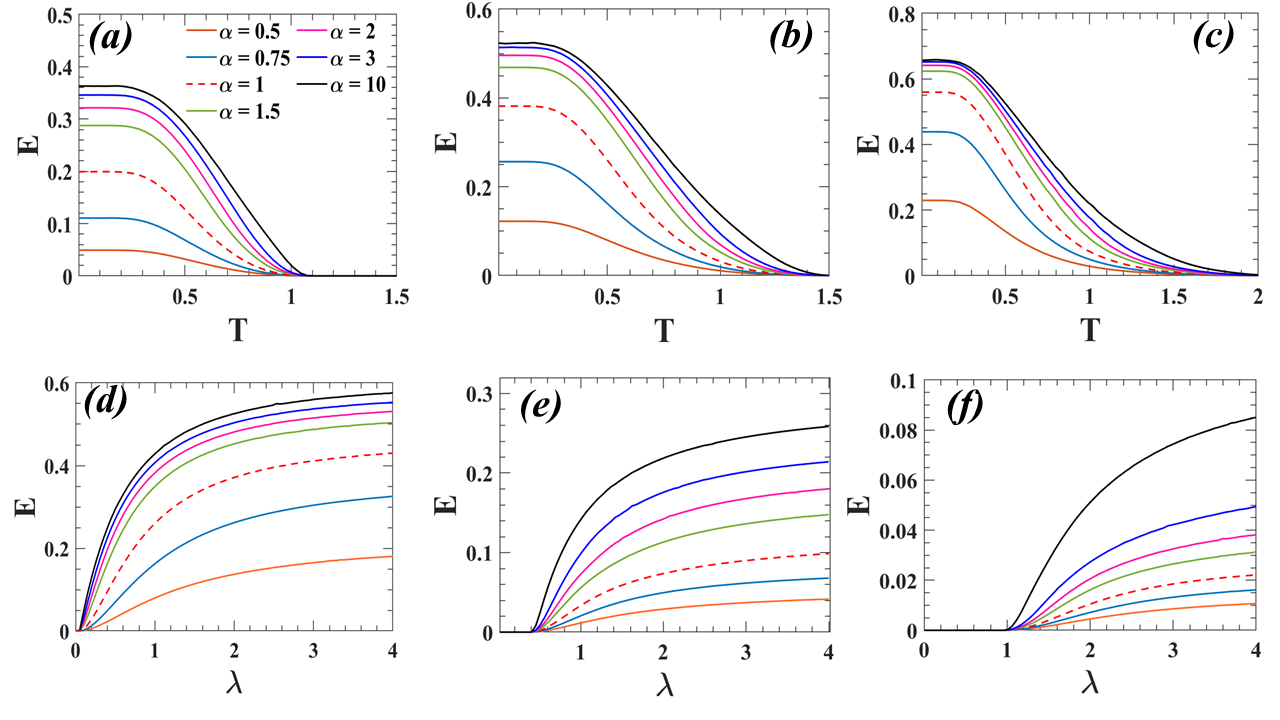}
\caption{Tripartite entanglement in the transverse-field Ising model calculated using the sandwiched R\'enyi relative entropy for 
various values of $\alpha$. The entanglement variation with temperature for (a) $\lambda = 0.5$, (b) $\lambda = 1$ and (c) $\lambda = 2$.  
Entanglement versus $\lambda.  $ is plotted for (d) $T = 0.5$, (e) $T = 1$ and (f) $T = 1.5$.}
\label{SandwichedIsing}
\end{figure}

The dependence of the entanglement in the transverse field Ising model with temperature $T$ is depicted in Fig. \ref{RenyiIsing} (a-c) and  
\ref{SandwichedIsing} (a-c) for different values of $\lambda$ and $\alpha$.  As usual, we find that the entanglement decreases as the temperature 
increases.  The entanglement at $T=0$ depends on the value of $\alpha$ and, moreover, the critical temperature $T_{c}$ is significantly shifted to 
larger values as $\lambda$ increases. 

An important observation can be made from a comparison of the plots for the traditional and sandwiched 
R\'enyi relative entropies of entanglement. We find that the entanglement calculated with the sandwiched R\'enyi relative is smaller than the 
corresponding traditional one.  This agrees very well with the Araki-Lieb-Thirring inequality for the two types of R\'enyi relative entropies 
satisfying \cite{inequality0,inequality1,inequality2,inequality3,inequality4}
\begin{equation}
    S_{\alpha}^{T}(\rho \| \sigma) \geq S_{\alpha}^{S}(\rho \| \sigma) \quad \forall  \alpha \in [0, 1) \cup (1, \infty).
    \label{inequalityRenyi}
\end{equation}
In Fig. \ref{RenyiIsing} (d-f) and  \ref{SandwichedIsing} (d-f), we plot the change of tripartite entanglement with the external field parameter 
$\lambda$ for different $\alpha$ and $T$.  In the regime of a very strong magnetic field, $B \rightarrow \infty$, i.e. $\left(\lambda =0\right)$, 
the spins are aligned with the external field and, consequently, the ground state is not entangled.  For a finite $B$, the thermal state is entangled.


\subsection{Monogamy of spin models}  
\label{Distribution}

The entanglement in the three-site Heisenberg model and the transverse field Ising model has been characterized using the R\'enyi relative entropy 
of entanglement in the previous two sections.  The measure is used to compute the total entanglement of the entire system, but it does not tell us how
the entanglement is distributed among its subsystems.  In this section, we look into the distribution of entanglement by examining the 
monogamy of the entanglement measures for the various models.  To do this, we find the quantity $M$ defined in Eq. \ref{MonogamyStatesRelation} 
for the Heisenberg models and the transverse field Ising model.  

\begin{figure}[hpbt] 
\includegraphics[width=0.90\columnwidth]{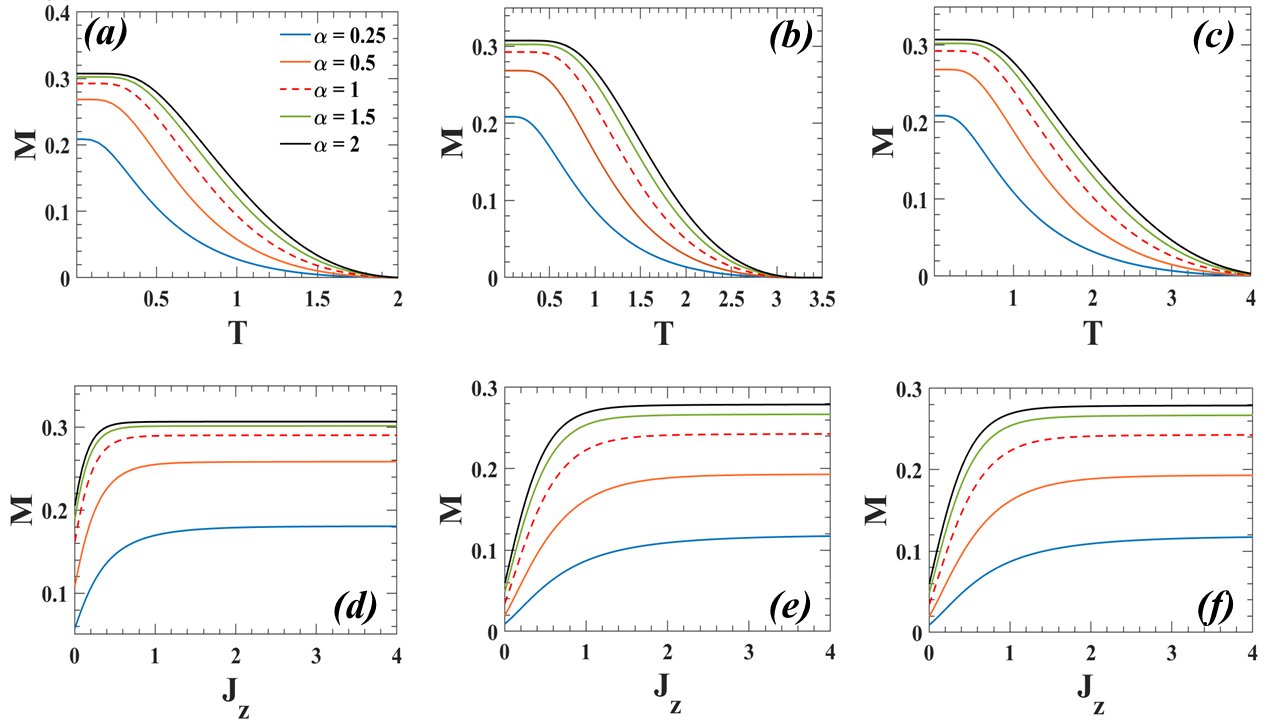}
\caption{Monogamy of entanglement of the XYZ model is calculated using traditional R\'enyi relative entropy for various values of $\alpha$.  
The variation of monogamy of entanglement with $T$ is shown for (a) $J_z = 0.2$, (b) $J_z = 1$ and (c) $J_z = 2$.   Plot of monogamy of 
entanglement versus $J_z$ is given for the temperature (d) $T = 0.5$, (e) $T = 1$ and (f) $T = 1.5$.  The values of $J_x$ and $J_y$ are fixed 
at  $0.8$ and $0.5$ respectively.}
\label{monoRenyiXYZ}
\end{figure}

\begin{figure}[hpbt] 
\includegraphics[width=0.90\columnwidth]{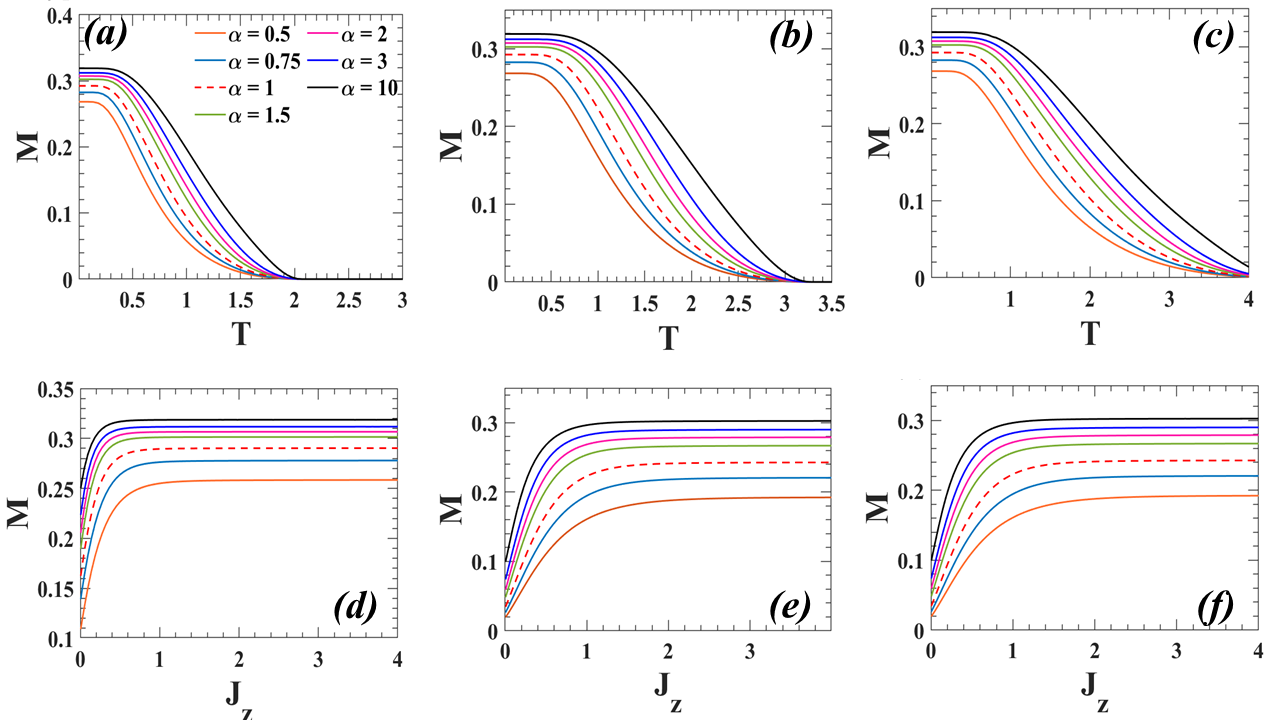}
\caption{Monogamy of entanglement of the XYZ model is calculated using sandwiched R\'enyi relative entropy for various values of $\alpha$.  
The variation of monogamy of entanglement with $T$ is shown for (a) $J_z = 0.2$, (b) $J_z = 1$ and (c) $J_z = 2$.   Plot of monogamy of 
entanglement versus $J_z$ is given for the temperature (d) $T = 0.5$, (e) $T = 1$ and (f) $T = 1.5$.  The values of $J_x$ and $J_y$ are fixed 
at  $0.8$ and $0.5$ respectively.}
\label{monoSandwichedXYZ}
\end{figure}

The variation of the monogamy of entanglement computed using the traditional R\'enyi relative entropy and the sandwiched R\'enyi relative entropy 
corresponding to the three-site XYZ model is presented in Fig. \ref{monoRenyiXYZ} and Fig. \ref{monoSandwichedXYZ}, respectively.  The plots 
Fig. \ref{monoRenyiXYZ} (a) - (c) and Fig. \ref{monoSandwichedXYZ} (a) - (c) illustrate the change in monogamy as a function of temperature 
for various values of $\alpha$.  In Fig. \ref{monoRenyiXYZ} (d)-(f) and Fig. \ref{monoSandwichedXYZ} (d)-(f) the variation of entanglement 
with the spin coupling parameter $J_{z}$ is shown for various values of $\alpha$.  The results establish that the states are monogamous at all temperatures.  

\begin{figure}[hpbt] 
\includegraphics[width=0.95\columnwidth]{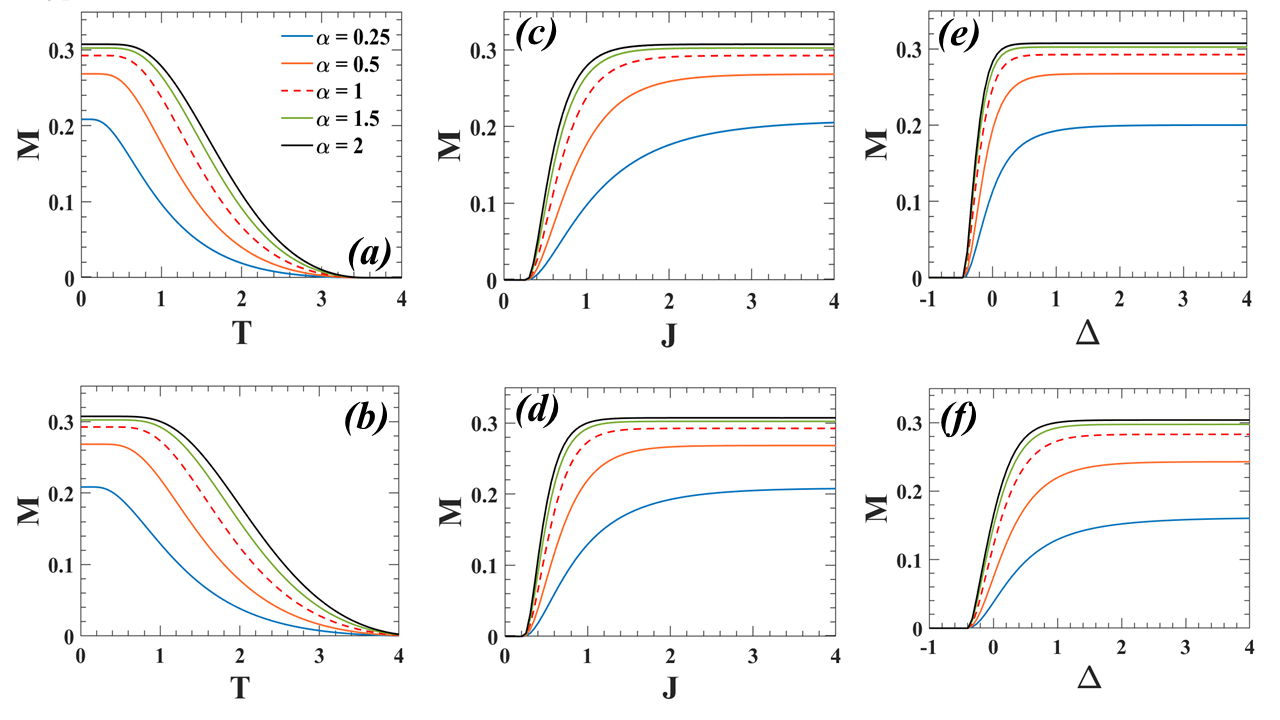}
\caption{The variation of monogamy of entanglement based on traditional R\'enyi relative entropy with temperature is shown for 
(a) $\Delta = 0.5$ and (b) $\Delta = 1$ for various values of $\alpha$, with $J = 1$. The change in monogamy of entanglement with respect to 
$J$ for (c) $\Delta = 0.5$ and (d) $\Delta = 1$ for different $\alpha$ keeping $T = 1$. The monogamy versus anisotropy $\Delta$ plot for (e) $T = 0.5$ and (f) $T = 1$ for various $\alpha$ keeping $J = 1$.}
\label{monoRenyiXXZ}
\end{figure}

\begin{figure}[hpbt] 
\includegraphics[width=0.95\columnwidth]{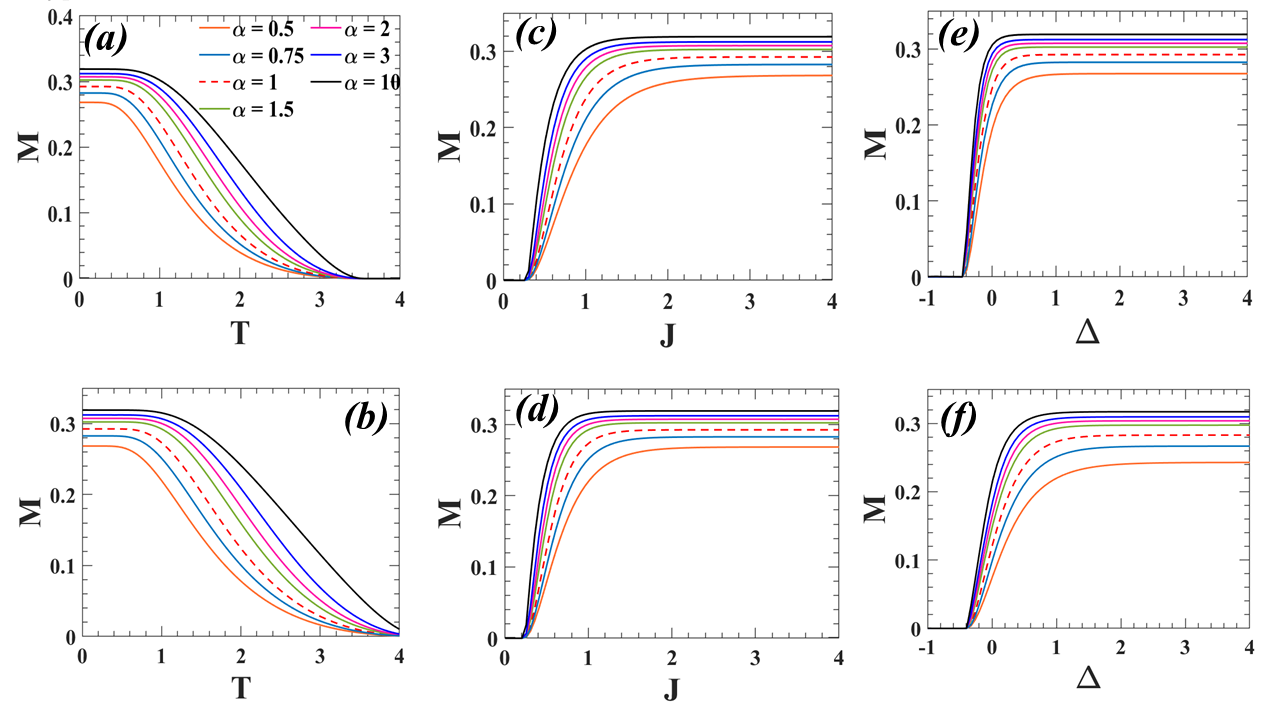}
\caption{The variation of monogamy of entanglement based on sandwiched R\'enyi relative entropy with temperature is shown for 
(a) $\Delta = 0.5$ and (b) $\Delta = 1$ for various values of $\alpha$, with $J = 1$. The change in monogamy of entanglement with 
respect to $J$ for (c) $\Delta = 0.5$ and (d) $\Delta = 1$ for different $\alpha$ keeping $T = 1$. The monogamy versus anisotropy 
$\Delta$ plot for (e) $T = 0.5$ and (f) $T = 1$ for various $\alpha$ keeping $J = 1$.}
\label{monoSandwichedXXZ}
\end{figure}

Next we examine the dynamics of monogamy of entanglement in the Heisenberg XXZ model.  In Fig. \ref{monoRenyiXXZ} and Fig. \ref{monoSandwichedXXZ},
we show the corresponding results for the traditional R\'enyi and the sandwiched R\'enyi relative entropies of entanglement.  The graphs in 
Fig. \ref{monoRenyiXXZ} (a) and (b) and Figs. \ref{monoSandwichedXXZ} (a) and (b)
show the variation of monogamy with respect to the temperature $T$.  In Fig. \ref{monoRenyiXXZ} (c) and (d) and Fig. \ref{monoSandwichedXXZ} 
(c) and (d), the monogamous nature is verified by varying the spin-spin coupling parameter for various values of $\alpha$.  Finally, the 
variation of entanglement with respect to the anisotropy parameter $\Delta$ is given in Fig. \ref{monoRenyiXXZ} (e) and (f) and 
Fig. \ref{monoSandwichedXXZ} (e) and (f) for the traditional and sandwiched R\'enyi relative entropy.  Here, we observe again that the states 
remain monogamous when the control parameters are varied.  

\begin{figure}[hpbt] 
\includegraphics[width=0.95\columnwidth]{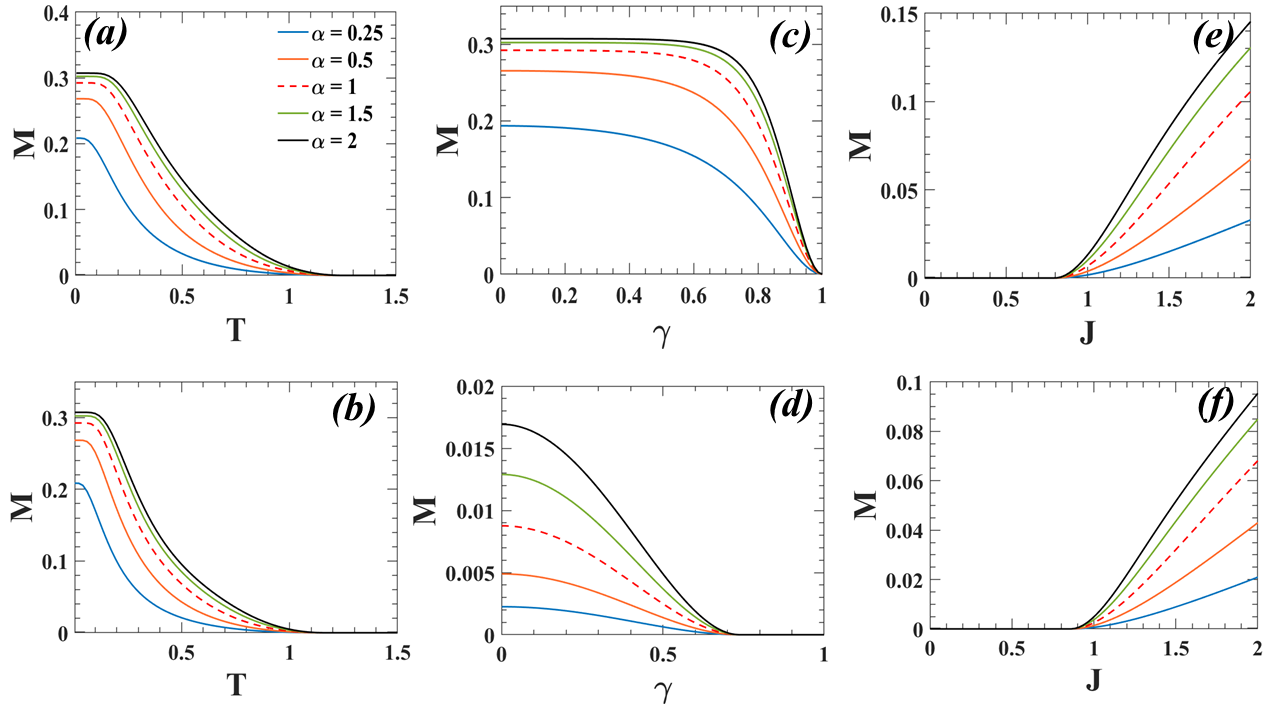}
\caption{The monogamy variation with temperature is shown for the traditional R\'enyi relative entropy for different values of $\alpha$.   
The change in monogamy is given as a function of temperature $T$ for (a) $\gamma = 0.25$ and (b) $\gamma = 0.50$ maintaining $J = 1$.  The 
dependence of monogamy with respect to $\gamma$ for (c) $T = 0.1$ and (d) $T = 1$ for $J = 1$.  Variation of monogamy with $J$ is given for (e) $\gamma = 
0.25$ and (f) $\gamma = 0.5$ with $T = 1$.}
\label{monoRenyiXY}
\end{figure}

\begin{figure}[hpbt] 
\includegraphics[width=0.95\columnwidth]{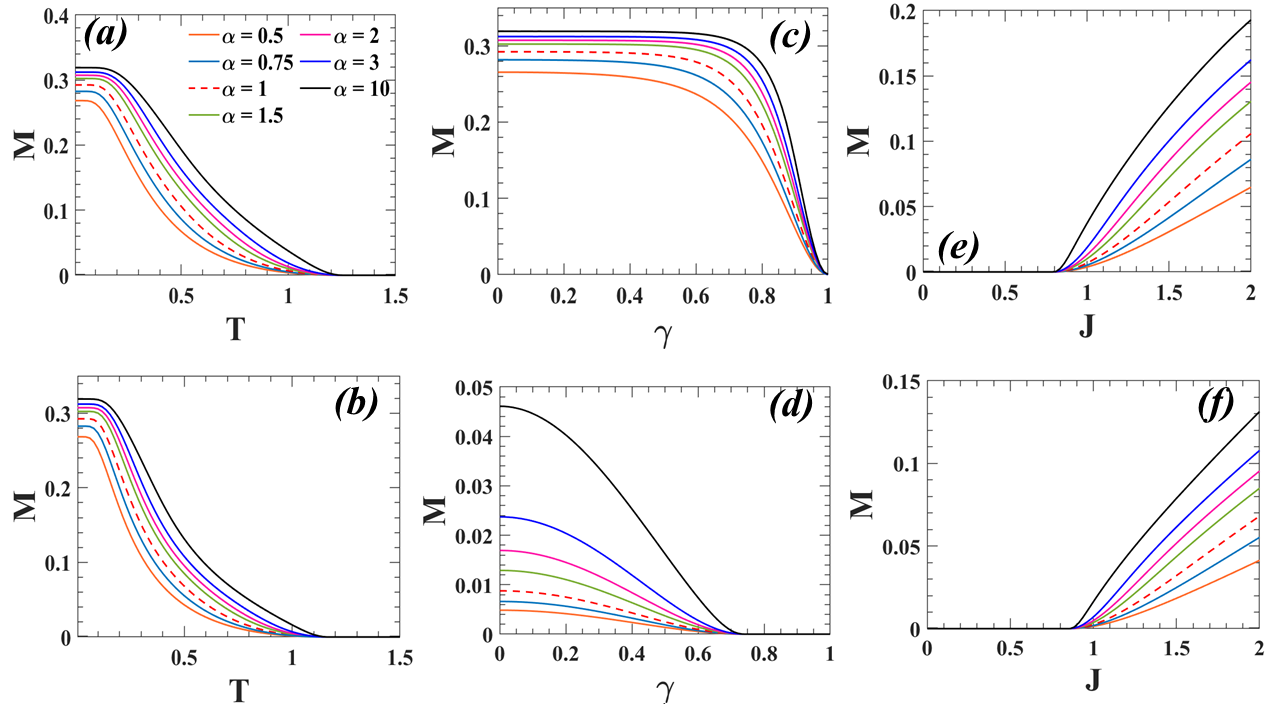}
\caption{The monogamy variation with temperature is shown for the sandwiched R\'enyi relative entropy for different values of $\alpha$.   
The change in monogamy is given as a function of temperature $T$ for (a) $\gamma = 0.25$ and (b) $\gamma = 0.50$ maintaining $J = 1$.  The 
dependence of monogamy with respect to $\gamma$ for (c) $T = 0.1$ and (d) $T = 1$ for $J = 1$.  Variation of monogamy with $J$ is given for (e) $\gamma = 
0.25$ and (f) $\gamma = 0.5$ with $T = 1$.}
\label{monoSandwichedXY}
\end{figure}

The monogamy of entanglement for the XY model calculated using 
the R\'enyi and sandwiched R\'enyi relative entropy are shown in Figs. \ref{monoRenyiXY} and \ref{monoSandwichedXY}, respectively.  
In Fig. \ref{monoRenyiXY} (a) and (b) as well as Fig. \ref{monoSandwichedXY} (a) and (b), we describe the variation of monogamy of 
entanglement as a function of temperature.  We observe that the state is monogamous for all temperatures. To understand the dependence on anisotropy, we also look at its effect on monogamy in Fig. \ref{monoRenyiXY} (c)  and (d), as well as Fig. \ref{monoSandwichedXY} 
(c) and (d) for the traditional R\'enyi and sandwiched R\'enyi relative entropy of entanglement, respectively.  Finally, in 
Fig. \ref{monoRenyiXY} (e) and (f) and in \ref{monoSandwichedXY} (e) and (f), the variation of monogamy with respect to the spin coupling 
parameter $J$ is discussed.  Here again, we find that the state is always monogamous.  Thus we observe that for the Heisenberg models, the 
entanglement is distributed in monogamous way, and it remains so until the state completely decoheres.  

\begin{figure}[hpbt] 
\includegraphics[width=0.95\columnwidth]{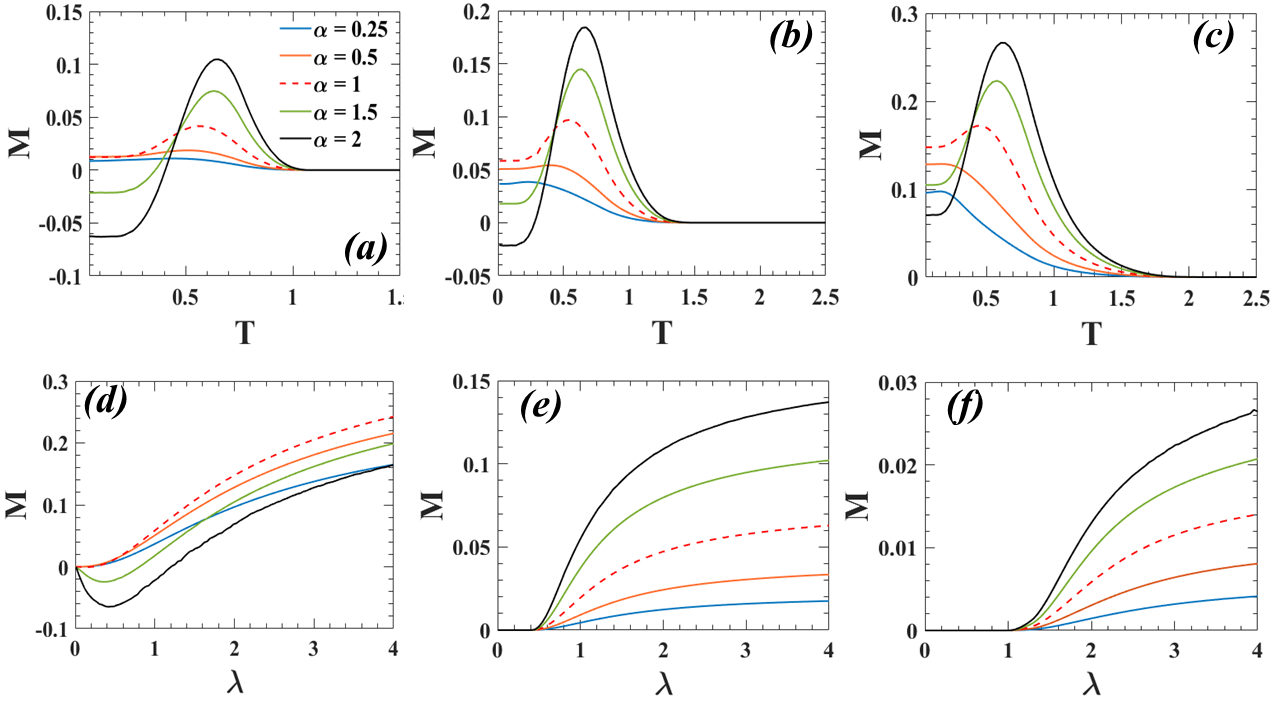}
\caption{Monogamy of entanglement of 1D transverse-field Ising model is computed using traditional R\'enyi relative entropy for several values 
of $\alpha$.  The variation of monogamy with temperature for (a) $\lambda = 0.5$, (b) $\lambda = 1$ and (c) $\lambda = 2$.  The change in monogamy 
with the magnetic field strength $\lambda$ for (d) $T = 0.5$, (e) $T = 1$ and (f) $T = 1.5$.  }
\label{monoRenyiIsing}
\end{figure}

\begin{figure}[hpbt] 
\includegraphics[width=0.95\columnwidth]{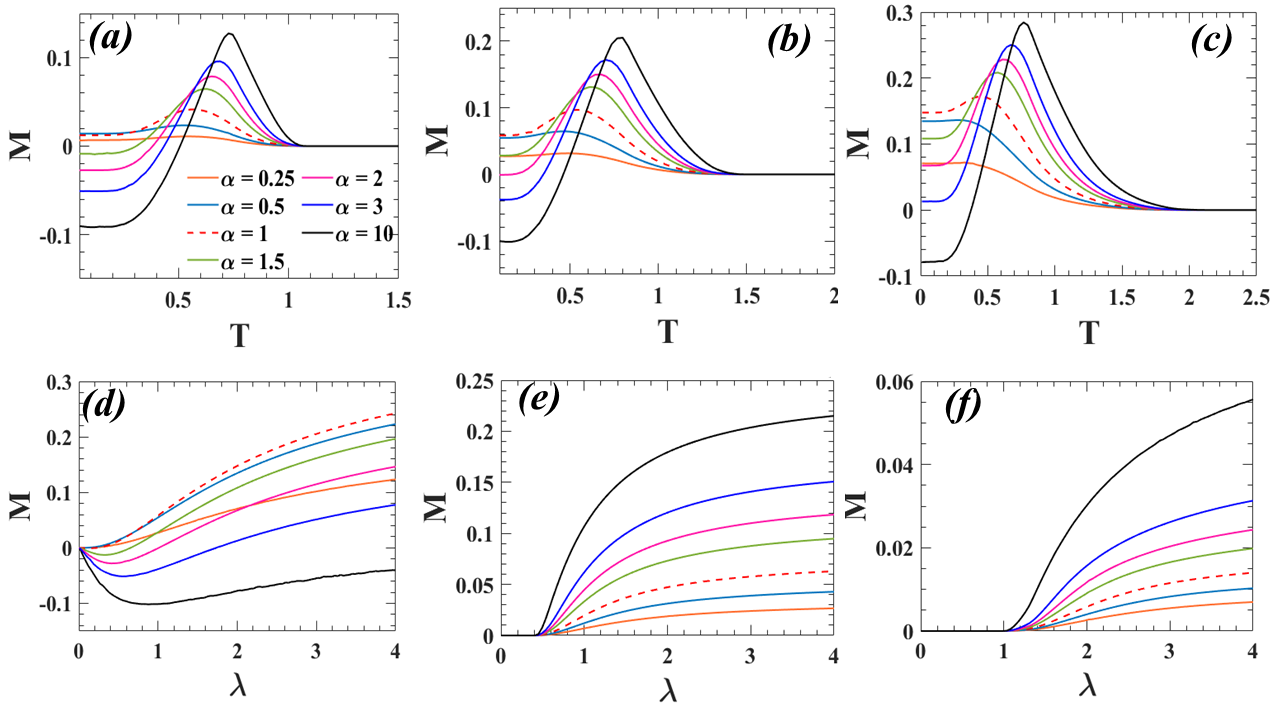}
\caption{Monogamy of entanglement of 1D transverse-field Ising model is computed using sandwiched R\'enyi relative entropy for several values 
of $\alpha$.  The variation of monogamy with temperature for (a) $\lambda = 0.5$, (b) $\lambda = 1$ and (c) $\lambda = 2$.  The change in monogamy 
with the magnetic field strength $\lambda$ for (d) $T = 0.5$, (e) $T = 1$ and (f) $T = 1.5$. }
\label{monoSandwichedIsing}
\end{figure}

All results shown so far have shown that the states are monogamous for all parameters. The transverse-field Ising model gives contrasting results, as shown in  Fig. \ref{monoRenyiIsing} and Fig. \ref{monoSandwichedIsing} for the traditional 
R\'enyi relative entropy and the sandwiched R\'enyi relative entropy, respectively.  The variation of the monogamy with temperature 
is shown in Fig \ref{monoRenyiIsing} (a) - (c) for the traditional R\'enyi relative entropy for different values of the 
spin-spin coupling parameter $\lambda$. When $\lambda = 0.5$ and $\lambda = 1$ as shown in Fig. \ref{monoRenyiIsing} (a) and 
Fig. \ref{monoRenyiIsing} (b) respectively, the initial state is polygamous and with increasing $\lambda$ it switches over to becoming monogamous. 
In Fig. \ref{monoRenyiIsing} (c) where $\lambda = 2$, the entanglement distribution is completely monogamous.  Since the parameter $\lambda$ is 
inversely proportional to the magnetic field $B$ ($\lambda = J/B$), we might think that for a large field the system exhibits a switch 
from polygamous to monogamous nature.  But this is not a conclusive observation because the R\'enyi relative entanglement measure can be used 
only up to $\alpha = 2$.  For a complete analysis considering large values of $\alpha$, we study the monogamy as a function of temperature $T$
for the sandwiched R\'enyi relative entropy in which the R\'enyi parameter is $0.5 \leq \alpha \leq \infty$.  The plots in 
Fig.\ref{monoSandwichedIsing} (a) and (b) confirm the switching action in monogamy of entanglement.  Interestingly in Fig.\ref{monoSandwichedIsing}
(c) we observe the transition from a polygamous to a monogamous nature in the quantum state for values of $\alpha > 2$. Thus, we find that it is 
important to use both the traditional and sandwiched R\'enyi relative entropy to investigate any quantum phenomena and consider all possible 
values of $\alpha$. 

We now give some insight into why the transverse-field Ising model yields polygamous states.  The ground state of 
the transverse-field Ising model 
\begin{align}
|\psi_{\text{TFI}} \rangle = \sin \phi |000 \rangle + \tfrac{1}{\sqrt{3}} \cos \phi(|011 \rangle + |101 \rangle + |110 \rangle)  \,.
\end{align}
When $\phi = 0$ the ground state is  $ \tfrac{1}{\sqrt{3}} (|011 \rangle + |101 \rangle + |110 \rangle) $ which is a symmetric $\overline{W}$ state 
which is a monogamous state for the  R\'enyi relative entropy, as discussed in Sec. \ref{MonogamyStates}. 

Tracing out one of the qubits, we obtain the density matrix
\begin{align}
\rho_{12} & = \text{Tr}_3 |\psi_{\text{TFI}} \rangle \langle \psi_{\text{TFI}} | \nonumber \\
& = (\sin^2 \phi + \tfrac{1}{3} \cos^2 \phi ) | v_1 \rangle \langle v_1 | 
+ \tfrac{2}{3} \cos^2 \phi | v_2 \rangle \langle v_2 | \,,
\end{align}
where the eigenstates are 
\begin{eqnarray}
   |v_{1} \rangle & = &  \frac{\sin \phi |00\rangle +  \frac{1}{\sqrt{3}} \cos \phi |11\rangle} 
                        {\sqrt{\sin^{2} \phi + \frac{1}{2} \cos^{2} \phi}} \\
  |v_{2} \rangle & = & \tfrac{1}{\sqrt{2}} \left(|01 \rangle + |10 \rangle \right)  .
\end{eqnarray}
The state $|v_{1} \rangle$ is partially entangled and $|v_{2} \rangle$ is a maximally entangled state.  For $\phi \rightarrow 0$, the partially entangled 
state $ | v_{1} \rangle $ becomes separable and the entanglement is only from the $| v_{2} \rangle $ and in such situation, the bipartite entanglements $E_{1:2}$
and $E_{1:3}$ are very low.  Consequently we have $E_{1:23} > E_{1:2} + E_{1:3}$ and hence the state exhibits monogamy.  When $\phi \neq 0$
we have situations where the combined value of $E_{1:2}$ and $E_{1:3}$ is greater than $E_{1:23}$ resulting in $E_{1:23} < E_{1:2} + E_{1:3}$
which leads to a polygamous distribution of entanglement.  This is a situation that is similar to the star state that was examined in Sec. 
\ref{MonogamyStates}.   
Further, the large values of $\alpha$ tend to give numerically larger values of entanglement as seen in  Secs. \ref{XYZ} and \ref{planar}, which can help in 
the detection of the switch in entanglement distribution.  
In contrast, if we use the Kullback–Leibler relative entropy, i.e., when $\alpha = 1$, such a switch in entanglement is not observed.


\section{Discussion}\label{Conclusions}
\hspace{1.5em}

We have studied the multipartite entanglement measure based on R\'enyi relative entropy and its distribution in several types of tripartite states.  In addition to standard tripartite states such as the GHZ, W, and star states, we examined the thermal states of the spin Heisenberg XYZ and XXZ models as well as in the planar spin chains such as the XY model and 
the transverse field Ising model.    
There are two distinct generalizations of the 
R\'enyi entropy, namely the {\it(i)} the traditional R\'enyi relative entropy and {\it(ii)} the sandwiched R\'enyi relative entropy. 
The traditional version of the R\'enyi relative entropy
is convex in the range  $\alpha \in [0,2]$ and the sandwiched entropy is convex in the region $\alpha \in [\frac{1}{2},\infty)$. In our study, 
we use both the traditional and sandwiched relative entropy to measure the entanglement within the regions where these measures are convex.  
The traditional and sandwiched R\'enyi relative entropies are also used to measure the monogamy of entanglement of various tripartite states. 
We find that both the GHZ and W states are monogamous for all values of $\alpha$. 
This is somewhat unexpected since W states are typically considered to be bipartite entangled and, hence, should display polygamous behavior. 
Meanwhile, for large values of $\alpha$ the star states switch from monogamous to polygamous nature.  
We attributed this to the higher purity of the reduced states of star states when tracing out one of the peripheral qubits, in comparison to W states.  Both states have entangled bipartite states, but this gives larger values of the bipartite entanglement in the case of R\'enyi relative entropies, which gives rise to the polygamous nature.  

We have given an in-depth analysis of the variation of entanglement in the spin models with different control variables like the temperature, the spin-spin interaction parameter, the anisotropy parameter, and the external magnetic field strength.  For both the XYZ and XXZ models, the entanglement 
decreases with temperature due to thermal decoherence.  The entanglement in the XYZ model increases with the interaction parameter $J_z$
for all values of $\alpha$ and temperature.  In the case of XXZ model, we characterize the entanglement with respect to the coupling 
strength $J$, and the anisotropy parameter $\Delta$ and find that the entanglement increases and attains a maximal value.  This 
maximal value is dependent on the R\'enyi $ \alpha $-parameter.  From the plots corresponding to the XYZ and the XXZ model, we notice 
that the entanglement in these models increases with  $\alpha$. 
We note that earlier studies have failed to detect entanglement \cite{ThermalXXZ} in all regions of the three site XXZ model, while the 
R\'enyi relative entropy has been successful in its detection.  The reason for this is that in earlier works bipartite measures 
to detect multipartite entanglement were used, and if the quantum state is of a GHZ form, the bipartite measures fail and hence multipartite measures are needed.

The qualitative behavior of entanglement with respect to different control variables has also been investigated for the Heisenberg 
XY model and the transverse field Ising model.  In both the models, the entanglement decreases with temperature as expected.  
With respect to the anisotropy parameter, the entanglement in the XY model decreases and the maximal entanglement is present at 
$\gamma = 0$, i.e., zero anisotropy which is the symmetric XX model.  Finally, we analyzed the variation of entanglement in the 
transverse field Ising model with the external magnetic field.  We find that the entanglement increases monotonically
and saturates to a finite value which depends on $\alpha$, the R\'enyi generalization parameter.  We also observe that the entanglement
calculated using the traditional R\'enyi relative entropy is higher than the sandwiched R\'enyi relative entropy and thus it obeys 
the Araki-Lieb-Thirring inequality.

Finally, to examine the distribution of entanglement we have analyzed the monogamy for all the spin models.  From our results
we find that the entanglement remains monogamous for the XYZ, XXZ, and XY models for all the control parameters under consideration. 
This implies that the fundamental nature of the entanglement distribution in Heisenberg models does not change when the control parameters are varied.  However,
in the transverse field Ising model, the entanglement distribution changes with the external parameters such as temperature $T$, 
the strength of the magnetic field $\lambda$, and the R\'enyi parameter $\alpha$.  From the results, we find that the entanglement which is 
initially polygamous changes to monogamous when the control parameters are increased.  This change in entanglement distribution 
can be attributed to the nature of the ground state of the transverse Ising model, 
which has a similar nature to star states in terms of the purity of the state after tracing out a qubit.  The bipartite reduced states of such
asymmetric states are a mixture of partially entangled and completely entangled states.  In the symmetric cases, the bipartite reduced 
states are a mixture of a separable state and a completely entangled state.  Consequently, the bipartite entanglement in the asymmetric
states are higher, and hence the state is polygamous.  When the control parameters are varied, the quantum state changes to a symmetric 
state, and hence the distribution switches to a monogamous nature.  Thus, the R\'enyi relative entropy of 
entanglement can highlight the asymmetry in the quantum state through the violation of monogamy inequality.  Further, the nature of 
switching between different entanglement distributions is a very interesting feature and has also been observed in the context of 
quantum coherence distribution in \cite{Chandrashekar,radhakrishnan2019basis}.  The existence of only two possible states of entanglement distribution 
and the ability to switch between using external controls can help in realizing switches based on quantum features.

\section*{Methods}

\subsection*{Density matrix of XYZ model}
\label{app:xyz}

The eigenvalues of the Hamiltonian in Eq. \eqref{HXYZ} are
\begin{align}
   E_0 &= E_7 = J_x+J_y+J_z+ \eta\,, \nonumber \\
    E_1 &= E_2 = E_4 = E_5 = -\left(J_x+J_y+J_z\right), \nonumber \\
    E_3 &= E_6 = J_x+J_y+J_z- \eta,
    \label{eigenvaluesXYZ}
\end{align}
and their respective eigenvectors are expressed in the standard basis as
\begin{align}
    \ket{\psi_{0}} &= \cos \phi_{0} \ket{\downarrow \downarrow \downarrow}+ \tfrac{1}{\sqrt{3}}\sin\phi_{0}\left(\ket{\uparrow \uparrow \downarrow}+\ket{\uparrow\downarrow\uparrow}+\ket{\downarrow\uparrow\uparrow}\right), \nonumber \\
    \ket{\psi_{7}} &= \cos \phi_{1} \ket{ \uparrow \uparrow \uparrow } + 
    \tfrac{1}{\sqrt{3}}\sin\phi_{1}\left(\ket{\downarrow\downarrow\uparrow}+\ket{\downarrow\uparrow\downarrow}+\ket{\uparrow\downarrow\downarrow}\right), \nonumber \\
    \ket{\psi_3} &= -\sin \phi_0 \ket{\downarrow \downarrow \downarrow} + 
    \tfrac{1}{\sqrt{3}}\cos\phi_0\left(\ket{\uparrow \uparrow \downarrow}+\ket{\uparrow\downarrow\uparrow}+\ket{\downarrow\uparrow\uparrow}\right), \nonumber \\
    \ket{\psi_6} &= -\sin \phi_1 \ket{ \uparrow \uparrow \uparrow } + 
    \tfrac{1}{\sqrt{3}}\cos\phi_1\left(\ket{\downarrow\downarrow\uparrow}+\ket{\downarrow\uparrow\downarrow}+\ket{\uparrow\downarrow\downarrow}\right), \nonumber \\
    \ket{\psi_1} &= \tfrac{1}{\sqrt{3}}\left(q\ket{\downarrow\downarrow\uparrow}+q^2\ket{\downarrow\uparrow\downarrow}+\ket{\uparrow\downarrow\downarrow}\right), \nonumber \\
    \ket{\psi_2} &= \tfrac{1}{\sqrt{3}}\left(q^2\ket{\downarrow\downarrow\uparrow}+q\ket{\downarrow\uparrow\downarrow}+\ket{\uparrow\downarrow\downarrow}\right), \nonumber \\
    \ket{\psi_4} &= \tfrac{1}{\sqrt{3}}\left(q\ket{\uparrow \uparrow \downarrow}+q^2\ket{\uparrow\downarrow\uparrow}+\ket{\downarrow\uparrow\uparrow}\right), \nonumber \\
    \ket{\psi_5} &= \tfrac{1}{\sqrt{3}}\left(q^2\ket{\uparrow \uparrow \downarrow}+q\ket{\uparrow\downarrow\uparrow}+\ket{\downarrow\uparrow\uparrow}\right). 
    \label{psiXY}
\end{align}
where $q=\exp\left({\frac{i2\pi}{3}}\right)$, $\eta=\sqrt{3\left(J_x-J_y\right)^2+\left(\left(J_x+J_y\right)-2J_z\right)^2}$ and
\begin{equation}
    \phi_{0}= \arctan\left(\frac{\sqrt{3} \left(J_x-J_y\right)}{2J_z-\left(J_x+J_y\right)+\eta} \right),\quad \phi_{1}=\arctan\left(\frac{\left(J_x+J_y\right)-2J_z+\eta}{\sqrt{3} \left(J_x-J_y\right)} \right).
\end{equation}
Once the system attains thermodynamical equilibrium with a thermal reservoir at temperature $T$, it can be described by the 
density matrix  $\rho\left(T\right)=\exp\left(-\beta H\right)/{Z}$, where $Z= \text{Tr} \left\{\exp\left(-\beta H\right)\right\}$ 
is the partition function of the system, and  $\beta= 1/k_{B} T$ with $k_{B}$ being the Boltzmann’s constant. For convenience, we will set $k=1$ 
throughout our discussion.

In the standard basis $\left\{\ket{\downarrow \downarrow \downarrow },
\ket{\downarrow \downarrow \uparrow},
\ket{\downarrow \uparrow \downarrow},
\ket{ \downarrow \uparrow \uparrow},
\ket{\uparrow \downarrow \downarrow},
\ket{\uparrow \downarrow \uparrow},
\ket{\uparrow \uparrow \downarrow},
\ket{ \uparrow \uparrow \uparrow }\right\}$, the thermal density matrix of the three-qubit XYZ model is
\begin{equation}
    \rho\left(T\right) = \frac{1}{Z}\begin{pmatrix}
    u & 0 & 0 & q_1 & 0 & q_1 & q_1 & 0 \\
    0 & w_1 & y_1 & 0 & y_1 & 0 & 0 & q_2 \\
    0 & y_1 & w_1 & 0 & y_1 & 0 & 0 & q_2 \\
    q_1 & 0 & 0 & w_2 & 0 & y_2 & y_2 & 0 \\
    0 & y_1 & y_1 & 0 & w_1 & 0 & 0 & q_2 \\
    q_1 & 0 & 0 & y_2 & 0 & w_2 & y_2 & 0\\
    q_1 & 0 & 0 & y_2 & 0 & y_2 & w_2 & 0 \\
    0 & q_2 & q_2 & 0 & q_2 & 0 & 0 & v \\
    \end{pmatrix}.
    \label{rhoXY}
\end{equation}
The matrix elements in Eq. \ref{rhoXY} are
\begin{align}
u&= x \left(e^{-\frac{\eta}{T^{\phantom{i}}}} \cos \phi_0^2+ e^{-\frac{\eta}{T^{\phantom{i}}}}  \sin \phi_0^2\right), \nonumber \\
v&=x \left(e^{-\frac{\eta}{T^{\phantom{i}}}} \cos \phi_1^2+ e^{-\frac{\eta}{T^{\phantom{i}}}} \sin \phi_1^2\right), \nonumber \\
w_1&=\tfrac{1}{3}x \left(2 x^{-2}+ e^{-\frac{\eta}{T^{\phantom{i}}}} \cos \phi_1^2+ e^{-\frac{\eta}{T^{\phantom{i}}}} \sin \phi_1^2\right), \nonumber \\
w_2&=\tfrac{1}{3} x\left(2 x^{-2}+ e^{-\frac{\eta}{T^{\phantom{i}}}} \cos \phi_0^2+ e^{-\frac{\eta}{T^{\phantom{i}}}} \sin \phi_0^2\right), \nonumber \\
y_1&=\tfrac{1}{3}x\left(-x^{-2}+ e^{-\frac{\eta}{T^{\phantom{i}}}} \cos \phi_1^2+ e^{-\frac{\eta}{T^{\phantom{i}}}}  \sin \phi_1^2\right), \nonumber \\
y_2&=\tfrac{1}{3}x\left(-x^{-2}+ e^{-\frac{\eta}{T^{\phantom{i}}}} \cos \phi_0^2+ e^{-\frac{\eta}{T^{\phantom{i}}}}  \sin \phi_0^2\right), \nonumber \\
q_1&=-\tfrac{2}{\sqrt{3}} x\cos \phi_0 \sin \phi_0 \sinh  \left(\frac{\eta}{T}\right), \nonumber \\
q_2&=-\tfrac{2}{\sqrt{3}} x\cos \phi_1 \sin \phi_1 \sinh  \left(\frac{\eta}{T}\right).
\end{align}
where $x=e^{-\frac{J_x+J_y+J_z}{T^{\phantom{i}}}}$. The analytic expression of the partition function is 
\begin{equation}
    Z=4x^{-1} + 4x \cosh \left(\frac{\eta}{T}\right).
\end{equation}

\subsection*{Density matrix of XXZ model}
\label{app:xxz}

The eigenvalues of the Hamiltonian in Eq. (\ref{HXXZ}) are
\begin{align}
   E_0 &= E_7 = 3J \Delta\,, \nonumber \\
    E_1 &= E_2 = E_4 = E_5 = -2J\left(\tfrac{1}{2}\Delta+1\right), \nonumber \\
    E_3 &= E_6 = -2J\left(\tfrac{1}{2}\Delta-2\right),
\end{align}
with the corresponding eigenstates
\begin{align}
    \ket{\psi_{0}} &= \ket{\downarrow \downarrow \downarrow}, \nonumber \\
    \ket{\psi_1} &= \tfrac{1}{\sqrt{3}}\left(q\ket{\downarrow\downarrow\uparrow}+q^2\ket{\downarrow\uparrow\downarrow}+\ket{\uparrow\downarrow\downarrow}\right), \nonumber \\
    \ket{\psi_2} &= \tfrac{1}{\sqrt{3}}\left(q^2\ket{\downarrow\downarrow\uparrow}+q\ket{\downarrow\uparrow\downarrow}+\ket{\uparrow\downarrow\downarrow}\right), \nonumber \\
    \ket{\psi_3} &=\tfrac{1}{\sqrt{3}}\left(\ket{\downarrow\downarrow\uparrow}+\ket{\downarrow\uparrow\downarrow}+\ket{\uparrow\downarrow\downarrow}\right), \nonumber \\
    \ket{\psi_4} &= \tfrac{1}{\sqrt{3}}\left(q\ket{\uparrow \uparrow \downarrow}+q^2\ket{\uparrow\downarrow\uparrow}+\ket{\downarrow\uparrow\uparrow}\right), \nonumber \\
    \ket{\psi_5} &= \tfrac{1}{\sqrt{3}}\left(q^2\ket{\uparrow \uparrow \downarrow}+q\ket{\uparrow\downarrow\uparrow}+\ket{\downarrow\uparrow\uparrow}\right),\nonumber \\ 
    \ket{\psi_6} &= \tfrac{1}{\sqrt{3}}\left(\ket{\uparrow \uparrow \downarrow}+\ket{\uparrow\downarrow\uparrow}+\ket{\downarrow\uparrow\uparrow}\right), \nonumber \\
    \ket{\psi_{7}} &= \ket{ \uparrow \uparrow \uparrow }. 
    \label{psiXXZ}
\end{align}
From the eigenvalues, the partition function of the system is $Z=2e^{\frac{-3J\Delta}{T^{\phantom{e}}}}+2 e^{\frac{J \Delta}{T^{\phantom{e}}}} 
\big(2e^{\frac{2J}{T^{\phantom{e}}}}+e^{-\frac{4J}{T^{\phantom{e}}}}\big)$.  Using the knowledge of the eigenvalues and eigenvectors of the system, we can construct 
the thermal density matrix of the system.  The entanglement in the density matrix is then calculated using the traditional R\'enyi
relative entropy of entanglement in Eq. (\ref{traditionalRenyi}) and the sandwiched R\'enyi relative entropy of entanglement in 
Eq. (\ref{SandwichedRenyi}).

\subsection*{Density matrix of the transverse-field Ising model}
\label{app:transverseising}

The eigenvalues and eigenvectors corresponding to the Hamiltonian are computed, which allows us to evaluate the thermal density matrix of the system, given by 
\begin{equation}
    \rho\left(T\right) = \frac{1}{Z}\begin{pmatrix}
    u & 0 & 0 & q_1 & 0 & q_1 & q_1 & 0 \\
    0 & w_1 & y_1 & 0 & y_1 & 0 & 0 & q_2 \\
    0 & y_1 & w_1 & 0 & y_1 & 0 & 0 & q_2 \\
    q_1 & 0 & 0 & w_2 & 0 & y_2 & y_2 & 0 \\
    0 & y_1 & y_1 & 0 & w_1 & 0 & 0 & q_2 \\
    q_1 & 0 & 0 & y_2 & 0 & w_2 & y_2 & 0\\
    q_1 & 0 & 0 & y_2 & 0 & y_2 & w_2 & 0 \\
    0 & q_2 & q_2 & 0 & q_2 & 0 & 0 & v \\
    \end{pmatrix}.
    \label{Isingdm}
\end{equation}
The elements of the density matrix are
\begin{align}
u&= e^{-\frac{J+B}{T^{\phantom{e}}}} \left(e^{-\frac{\eta_1}{T^{\phantom{i}}}} \cos \phi_0^2+ e^{\frac{\eta_1}{T^{\phantom{i}}}}  \sin \phi_0^2\right), \nonumber \\
v&=e^{-\frac{J-B}{T^{\phantom{i}}}} \left(e^{-\frac{\eta_2}{T^{\phantom{i}}}} \cos \phi_1^2+ e^{\frac{\eta_2}{T^{\phantom{i}}}}  \sin \phi_1^2\right), \nonumber \\
w_1&=\tfrac{1}{3}e^{-\frac{J-B}{T^{\phantom{i}}}} \left(2 e^{\frac{2\left(J-B\right)}{T^{\phantom{i}}}}+ e^{\frac{\eta_2}{T^{\phantom{i}}}}  \cos \phi_1^2+ e^{-\frac{\eta_2}{T^{\phantom{i}}}}  \sin \phi_1^2\right), \nonumber \\
w_2&=\tfrac{1}{3}e^{-\frac{J+B}{T^{\phantom{i}}}}\left(2 e^{\frac{2\left(J+B\right)}{T^{\phantom{i}}}}+ e^{\frac{\eta_1}{T^{\phantom{i}}}}  \cos \phi_0^2+ e^{-\frac{\eta_1}{T^{\phantom{i}}}}  \sin \phi_0^2\right), \nonumber \\
y_1&=\tfrac{1}{3}e^{-\frac{J-B}{T^{\phantom{i}}}} \left(- e^{\frac{2\left(J-B\right)}{T^{\phantom{i}}}}+ e^{\frac{\eta_2}{T^{\phantom{i}}}}  \cos \phi_1^2+ e^{-\frac{\eta_2}{T^{\phantom{i}}}}  \sin \phi_1^2\right), \nonumber \\
y_2&=\tfrac{1}{3}e^{-\frac{J+B}{T^{\phantom{i}}}}\left(- e^{\frac{2\left(J+B\right)}{T^{\phantom{i}}}}+ e^{\frac{\eta_1}{T^{\phantom{i}}}}  \cos \phi_0^2+ e^{-\frac{\eta_1}{T^{\phantom{i}}}}  \sin \phi_0^2\right),\nonumber \\
q_1&=-\tfrac{2}{\sqrt{3}}e^{-\frac{J+B}{T^{\phantom{i}}}} \cos \phi_0 \sin \phi_0 \sinh  \left(\frac{\eta_1}{T^{\phantom{i}}}\right), \nonumber \\
q_2&=-\tfrac{2}{\sqrt{3}}e^{-\frac{J-B}{T^{\phantom{i}}}} \cos \phi_1 \sin \phi_1 \sinh  \left(\frac{\eta_2}{T^{\phantom{i}}}\right).
\end{align}
where
\begin{equation}
    \eta_{1}= 2\sqrt{1-\lambda+\lambda^2},\quad  \eta_{2}=2\sqrt{1+\lambda+\lambda^2},
\end{equation}
and
\begin{equation}
    \phi_{0}= \arctan\left(\frac{\sqrt{3} \lambda}{2-\lambda +\eta_1} \right),\quad \phi_{1}=\arctan\left(\frac{2+\lambda+\eta_2}{\sqrt{3} \lambda} \right).
\end{equation}
The partition function in the system reads $Z=4e^{\frac{\lambda}{T^{\phantom{i}}}} \cosh \left(\frac{1}{T}\right) 
+ 2e^{-\frac{\lambda+1}{T^{\phantom{i}}}} \cosh \left(\frac{\eta_1}
{T}\right) + 2e^{-\frac{\lambda-1}{T^{\phantom{i}}}} \cosh \left(\frac{\eta_2}{T}\right)$.

\section*{Acknowledgements}
M.M., T. B., H.S. acknowledge support by Tamkeen under the NYU Abu Dhabi Research Institute grant CG008. We
acknowledge the computations performed at the NYU IT High Performance Computing services.  T. B. is supported by the National Natural Science Foundation of China (62071301); NYU-ECNU Institute of Physics at NYU Shanghai; the Joint Physics Research Institute Challenge Grant; the Science and Technology Commission of Shanghai Municipality (19XD1423000,22ZR1444600); the NYU Shanghai Boost Fund; the China Foreign Experts Program (G2021013002L); the NYU Shanghai Major-Grants Seed Fund; and the SMEC Scientific Research Innovation Project (2023ZKZD55).

\section*{Competing Interests}
The authors declare no competing interests.

\section*{Author Contributions}
MM and CR performed the calculations. CR conceived the work.  TB and HS guided the work.  All authors wrote the paper.

\bibliography{Marwa.bib}

%
%
%

\end{document}